\documentclass{article}
\usepackage{hchang}
\usepackage{hyperref}
\usepackage{comment}
\usepackage{tcolorbox}
\usepackage{color}
\usepackage{graphicx} 
  \setkeys{Gin}{width=\linewidth,totalheight=\textheight,keepaspectratio}
  \graphicspath{{graphics/}} 
\usepackage{booktabs}
\usepackage{array} 
\usepackage{units}    
\usepackage{multicol} 
\usepackage{lipsum}   
\usepackage{fancyvrb} 
  \fvset{fontsize=\normalsize}
\usepackage[capitalise]{cleveref}
\usepackage{rotating}
\usepackage{transparent}
\usepackage{multicol}
\usepackage{tikz}
\usetikzlibrary{calc,decorations.pathreplacing,shapes.geometric}
\usepackage{fullpage}
\usepackage{soul}
\usepackage{thmtools}
\usepackage{thm-restate}
\usepackage{algorithm}
\usepackage{algorithmicx}
\usepackage{algpseudocode}
\usepackage{arydshln}


\newtheorem{lemma}{Lemma}
\newtheorem{theorem}{Theorem}
\newtheorem{claim}{Claim}
\newtheorem{definition}{Definition}
\newtheorem{observation}{Observation}
\newtheorem{corollary}{Corollary}
\newtheorem{remark}{Remark}

\newtheorem{goal}{Goal}
\newtheorem{conjecture}{Conjecture}

\newcommand{\eps}{\epsilon}

\newcommand{\draw}{\in_{R}}

\newcommand{\real}{\mathbb{R}}
\newcommand{\opt}{\mathsf{Opt}}

\newcommand{\cP}{\mathcal{P}}
\newcommand{\cT}{\mathcal{T}}

\newcommand{\bbE}{\mathbb{E}}

\newcommand{\cD}{\mathcal{D}}

\newcommand{\lightgray}[1]{\textcolor{gray}{#1}}

\crefname{problem}{problem}{problems}
\Crefname{problem}{Problem}{Problems}
\crefformat{problem}{problem~(#2#1#3)}
\Crefformat{problem}{Problem~(#2#1#3)}

\renewcommand{\leq}{\leqslant}

\renewcommand{\geq}{\geqslant}
\renewcommand{\ge}{\geqslant}

\def\cB{\ensuremath{\mathcal{B}}}  
\def\cC{\ensuremath{\mathcal{C}}}
\def\cD{\ensuremath{\mathcal{D}}}
\def\cE{\ensuremath{\mathcal{E}}}

\def\cI{\ensuremath{\mathcal{I}}}

\def\cO{\ensuremath{\mathcal{O}}}
\def\cP{\ensuremath{\mathcal{P}}}

\def\cT{\ensuremath{\mathcal{T}}}

\def\cV{\ensuremath{\mathcal{V}}}

\def\bbC{\ensuremath{\mathbb{C}}}

\algtext*{EndIf}
\algtext*{EndFor}
\algtext*{EndWhile}
\algtext*{EndProcedure}

\def\opt{\ensuremath{\mathsf{Opt}}}
\def\torso{\ensuremath{\mathsf{Torso}}}
\def\apices{\ensuremath{\mathsf{apices}}}
\def\what{\widehat}
\def\frakH{\mathfrak{H}}

\newif\ifpaperreview
\paperreviewtrue

\title{Planar Graphs Embed into Graphs with Treewidth $O(\log n /\epsilon)$}

\author{
Hung Le\thanks{University of Massachusetts Amherst.  \texttt{hungle@cs.umass.edu}. Supported by the NSF CAREER award CCF-2237288 and the NSF grant CCF-2517033.}}
\date{}

\begin{document}

\maketitle

\begin{abstract}

We construct a stochastic embedding of $n$-vertex planar graphs into graphs of treewidth $O(\log n/\eps)$ and expected distortion $1+\eps$ for any parameter $\eps \in (0,1)$. This result yields (randomized) polynomial-time approximation schemes (PTASes) for a broad class of connectivity problems on planar graphs, including the $k$-MST problem. The treewidth bound is asymptotically optimal and settles a conjecture of Chang, Cohen-Addad, Conroy, Le, Ma. Pilipczuk, and Mi. Pilipczuk~\cite{CCCLPP25}.

The key idea is a new construction of a stochastic balanced cut of size $O(\log n/\eps)$ and expected distortion $(1+\eps/\log n)$ for every edge.  We construct the cuts via the minimax principle, which differs significantly from existing cut constructions based on a random clustering hierarchy.

Our embedding extends to minor-free graphs using the Robertson--Seymour decomposition, giving the same treewidth bound. Using this decomposition, we obtain the first (randomized) PTASes for many connectivity problems in minor-free graphs, including the Steiner tree problem and its prize-collecting variant. 

\end{abstract}

\paragraph{AI Disclosure.} The construction of the stochastic balanced cut, a central technical idea of this paper, is due to GPT-6 Astra. The author attempted to solve this problem several times with different ChatGPT models, starting with GPT-5.5, without success. Until very recently, GPT-6 Astra made a breakthrough after 26 prompts when the author directed it to examine the column model idea in the solution to the L1 embedding conjecture by OpenAI~\cite {O26}. Curiously, the column model idea ultimately did not make it into the final proof, which is surprisingly elementary and only uses known tools.  The author introduced other ideas, namely algorithmic minimax and the extension to minor-free graphs. The author aslo completely rewrote the paper, contextualized the results and new techniques, and attributed existing ideas to prior work (to the best of the author's knowledge). ChatGPT was also used to proofread the paper; otherwise, it is written by a human and for human readers.

\begingroup
\normalsize
\nolinenumbers
\setcounter{tocdepth}{2}
\setlength{\columnsep}{1.5em}
\setlength{\multicolsep}{6pt}
\vspace{30pt}
\begin{multicols}{2}[\noindent{\normalsize\bfseries\contentsname}]
\makeatletter
\@starttoc{toc}
\makeatother
\end{multicols}
\endgroup
\thispagestyle{empty}
\clearpage
\pagenumbering{arabic}

\section{Introduction}

Over the last few decades, research on designing polynomial-time approximation schemes (PTASes) for optimization problems on planar graphs has produced several powerful algorithmic frameworks that apply broadly to many problems.  A notable milestone is Baker's layering technique~\cite{B94}, which yields PTASes for many local problems on planar graphs, such as independent set, vertex cover, and dominating set, to name a few. These problems are local in the sense that feasibility can be checked by looking at each vertex's neighborhood. Baker's technique was then abstracted into a so-called \EMPH{deletion decomposition}~\cite{DHM10} and generalized to minor-free graphs~\cite{DHK05}. In a nutshell, the deletion decomposition says that for any integer parameter $k \geq 1$ one can partition the edges of the input (planar/minor-free) graphs into $k+1$ sets such that \underline{deleting} any one of the sets gives a graph of treewidth $O(k)$.

As powerful as they are, Baker's technique and the deletion decomposition cannot handle (non-local) \EMPH{connectivity problems} such as TSP, Steiner tree, and $k$-MST. This motivates the \EMPH{contraction decomposition} technique~\cite{K05,DHM10}: given any $k\geq 1$, one can partition the edges into $(k+1)$-sets such that \underline{contracting} any one set yields a graph of treewidth $O(k)$. However, contraction decomposition alone does not imply a PTAS: it needs a  \EMPH{spanner step}, where, given a connectivity problem $\Pi$ (e.g., TSP), construct a subgraph $S$ of $G$, called a \EMPH{$(1+\eps)$-spanner}, such that:
\begin{enumerate}
    \item The (value of the)  optimal solution for $\Pi$ in $S$ is within $(1+\eps)$-approximation of the (value of the) optimal solution for $\Pi$ in $G$, denoted by $\opt_G$. 
    \item The total length of $S$, denoted by $\ell(S)$, must be at most a constant times $\opt_G$:  $\ell(S) =f(\eps)\opt_G$, for some function $f$ that depends on $\eps$ only.
\end{enumerate}

Klein~\cite{K05} (implicitly) showed that, under a very mild condition\footnote{The condition is that $\Pi$ admits a polynomial-time algorithm or a PTAS in bounded-treewidth graphs. This condition holds for many problems due to Courcelle's theorem~\cite{C90}.}, if $\Pi$ admits a $(1+\eps)$-spanner on planar graphs, then one can design a PTAS for $\Pi$ by (1) constructing the spanner in polynomial time and (2) applying the contraction decomposition to the spanner.  This framework also extends to minor-free graphs~\cite{DHK11}.  Unfortunately, the spanner step is very ad hoc and problem-specific. 

Constructing good spanners for even very basic problems on planar graphs required substantial research effort, including subset TSP~\cite{K06}, Steiner tree~\cite{BKM09}, prize-collecting Steiner tree~\cite{BCEHKM11}, and Steiner forest~\cite{BHM11}. Each problem requires significantly new ideas to construct the spanner. Furthermore, among these problems, only spanners for TSP~\cite{BLW17} and Subset TSP~\cite{L20,CFKL20} have been generalized to minor-free graphs, also with completely different techniques. Designing a PTAS for Steiner tree in minor-free graphs remains an outstanding open problem; only a QPTAS is known~\cite{C22}. Given the ad hoc nature of the spanner construction, a central goal has become searching for a general-purpose technique.

\begin{goal}\label{goal:unified} Develop an algorithmic technique that gives PTASes for a broad range of connectivity problems in planar and minor-free graphs.
\end{goal}

One guiding problem toward \Cref{goal:unified} is the \EMPH{$k$-minimum spanning tree} ($k$-MST) problem: given a graph $G$ with edge lengths and an integer parameter $k\geq 1$, find a tree of minimum length that spans at least $k$ vertices. This problem has been studied extensively in network design~\cite{RSMRR96,AABV95,ARV99,Garg96,AR98,AK06} and admits a polynomial-time $2$-approximation in general graphs~\cite{Garg05}. In planar graphs, the difficulty lies in locating the near-optimal solution. For example, when $k = \sqrt{n}$, the optimal solution might use only a small fraction of the vertices, while it is hard to know which vertices to retain, so the spanner step in the contraction decomposition framework does not seem to work. Thus, research on this problem has avoided the spanner step and, in doing so, has opened two different paths to the solution, which are stepping stones toward \cref{goal:unified}.

The first technique, introduced by Cohen-Addad~\cite{C22}, constructs a nondeterministic split tree (NDST) for planar and minor-free graphs. Basically, it is a randomized decomposition of the input graph into a hierarchy of clusters such that a near-optimal solution of a problem $\Pi$ only crosses each cluster $O(\polylog(n))$ times. Thus, in quasi-polynomial time, one can enumerate all these crossing points and design a dynamic programming algorithm to find a $(1+\eps)$-approximate solution.  This technique is applicable to $k$-MST as well as all other connectivity problems mentioned above, yielding a QPTAS in minor-free graphs with running time $n^{O(\polylog(n))}$. Conceptually, the NDST technique is significant progress toward \Cref{goal:unified}. Yet, it is unclear if this technique could give a PTAS due to several technical obstacles, the most serious being the number of crossings, for which $\Omega(\log n)$ seems necessary. 

A second technique, pioneered by Cohen-Addad, Le, Ma. Pilipczuk, and Mi. Pilipczuk (CLPP), is stochastic embeddings into low-treewidth graphs~\cite{CLPP23} with expected distortion of $1+\eps$ for any $\eps \in (0,1)$. They showed that for planar graphs with polynomial aspect ratio, $\EMPH{$\Phi$} \coloneqq \frac{\max_{u,v}d_G(u,v)}{\min_{u\not= v}d_G(u,v)} = n^{O(1)}$, the treewidth of their embedding is $O(\log^{13}n/\eps)$. Their construction extends to minor-free graph with another factor of $\log(n)$ in the treewidth. This result again yields QPTASes for all the aforementioned connectivity problems in minor-free graphs and more~\cite{CLPP23}. Most importantly, their work provides a concrete path toward realizing \Cref{goal:unified} by further reducing treewidth. Specifically, \cite{CCCLPP25} refined the construction and reduced the treewidth to $O(\log^3 /\eps)$ for \underline{planar graphs}. They also conjectured that:

\begin{conjecture}[Conjecture 1.1~\cite{CCCLPP25}]\label{conj:main} Any $n$-vertex planar graph with edge lengths can be embedded with expected distortion $1+\eps$ into a distribution of graphs with treewidth $O(\log n/\eps)$ for any $\eps\in (0,1)$.
\end{conjecture}

The authors of~\cite{CCCLPP25} remarked that a (polynomial-time) resolution of \Cref{conj:main} yields a PTAS for the $k$-MST problem. More generally, such an embedding gives PTASes for a very broad class of problems that admit an algorithm (exact or PTASes) in treewidth-$\tau$ graphs in time $2^{O(\tau)}n^{O(1)}$, since when $\tau = O(\log n/\eps)$, the running time is polynomial for any fixed $\eps$. This class includes TSP, Steiner tree, its prize-collecting versions, the $k$-MST problem, $k$-stroll, and many other problems for which no PTAS is known on planar graphs. Thus, a polynomial-time embedding satisfying \Cref{conj:main} would achieve \Cref{goal:unified}.

Finally, we remark that there is not much wiggle room for achieving PTASes through low-treewidth embeddings. Any slightly larger treewidth bound, say $O(\log(n)\log^*(n)/\eps)$, would not be enough to give PTASes, while there is a lower bound of $\Omega(\log n)$~\cite{CG04} for even a constant distortion. One really needs an asymptotically optimal treewidth embedding to achieve \Cref{goal:unified}.  On a technical level, the authors of~\cite{CCCLPP25} have exhausted all possible improvements in the CLPP framework~\cite{CLPP23}: every step in the framework is already tight. They remarked~\cite{CCCLPP25} ``to further improve any logarithmic factor, one has to significantly deviate from the framework of [CLPP23]''.

\subsection{Our Results}

Our main result is a polynomial-time resolution of \Cref{conj:main}. 

\begin{restatable}{theorem}{mainthm}\label{thm:main}
Given an $n$-vertex planar graph $G = (V,E,\ell)$ and any parameter $\eps\in (0,1)$, there is a stochastic embedding of $G$ into a distribution $\cD$ of graphs with treewidth $O(\log n/\eps)$ and expected distortion $(1+\eps)$. Furthermore, the embedding can be sampled in polynomial time, and every graph $H$ in the support of $\cD$ has $V(H) = V(G)$. 
\end{restatable}

Not only does \Cref{thm:main} achieve an asymptotically optimal treewidth bound, but it also does not depend on the aspect ratio $\Phi$: previous embeddings~\cite{CCCLPP25,CLPP23} have $\polylog(\Phi)$ in the treewidth. Curiously, our embedding in \Cref{thm:main} does not require Steiner points. That is, the vertex set of the host graph $H$ is the same as that of $G$. This is technically more convenient than previous embeddings which use Steiner points, as each edge of $H$ can be mapped exactly to the shortest path between its endpoints in $G$.

A key new technical idea in the proof of \Cref{thm:main} is a new construction of a \EMPH{stochastic balanced cut}. Stochastic balanced cuts were introduced in CLPP embedding~\cite{CLPP23} and further refined in~\cite{CCCLPP25}. However, the cuts constructed in these papers are based on a clustering hierarchy, which is not only complicated but also yields suboptimal guarantees. The construction in this paper is much simpler, has optimal guarantees, and does not rely on such a clustering hierarchy. We provide more detail on the new construction in \Cref{subsec:tech-ideas}.   

The new balanced cut construction for planar graphs extends to minor-free graphs using Robertson--Seymour type decomposition.  Previous balanced cuts~\cite{CLPP23} also extend to minor-free graphs, also via the Robertson--Seymour decomposition, but lose another factor of $\log(n)$ in the treewidth. Our balanced cut does not have such a loss, and therefore, we obtain an optimal treewidth bound in minor-free graphs as well.

\begin{restatable}{theorem}{minorembthm}\label{thm:minor-tw} Fix an integer $h\geq 1$. Given an $n$-vertex $K_h$-minor-free graph $G = (V,E,\ell)$ and any parameter $\eps\in (0,1)$, there is a stochastic embedding of $G$ into a distribution $\cD$ of graphs with treewidth $O_h(\log n/\eps)$ and expected distortion $(1+\eps)$, where the $O_h$ notation hides a dependency on the minor.   The embedding can be sampled in polynomial time, and every graph $H$ in the support of $\cD$ has $V(H) = V(G)$. 
\end{restatable}

As we discussed above, the embeddings in \Cref{thm:main} and \Cref{thm:minor-tw} give (randomized) PTASes for a broad range of connectivity problems. Notably, we obtain the first randomized PTAS for $k$-MST in planar and minor-free graphs, and the first randomized PTAS for Steiner tree in minor-free graphs. For other problems, readers are referred to \Cref{sec:apps} for precise definitions.

\begin{restatable}{theorem}{ptases}\label{thm:PTASes}
Steiner tree, prize-collecting Steiner tree, $k$-MST, $k$-TSP, and  $k$-Stroll  admit randomized PTASes on planar and minor-free graphs.
\end{restatable}

\subsection{Key Technical Ideas}\label{subsec:tech-ideas}

Let $G = (V,E,\ell)$ be the input graph with non-negative edge length function $\ell$.  A key technical concept in our embedding algorithm is a new construction of a stochastic balanced cut with low routing cost. This notion was introduced in the CLPP paper~\cite{CLPP23}, though the precise definition differs slightly from ours due to the technical nature of their work.  In this overview, we will give a simplified definition of the balanced cuts. In the actual proof, we need to account for the recursion in the embedding, so the precise definition must include boundary vertices (passed down from earlier recursive steps), which makes it slightly more complicated.

\begin{figure}[t]
  \centering
  \resizebox{0.9\linewidth}{!}{%
  \begin{tikzpicture}[
    x=1cm, y=1cm, line cap=round, line join=round,
    graph edge/.style={draw=edgeink,line width=.8pt},
    vertex/.style={
      circle,inner sep=0pt,minimum size=5.4pt,
      fill=vertexink,draw=white,line width=.5pt
    },
    center vertex/.style={
      rectangle,inner sep=0pt,minimum size=7pt,
      fill=white,draw=black,line width=1pt
    },
    cluster/.style={
      fill=yellow!13,line width=.7pt,dash pattern=on 2pt off 1.7pt
    },
    vertex label/.style={font=\small,inner sep=.5pt},
    route/.style={
      line width=2pt,preaction={draw=white,line width=3.2pt}
    }
  ]
    \definecolor{toCenter}{HTML}{076CA3}
    \definecolor{fromCenter}{HTML}{B94B18}
    \definecolor{outlineC}{HTML}{BF528C}
    \definecolor{outlineD}{HTML}{BF792D}

    \foreach \panel/\offset in {0/0cm,1/7.36cm}{
      \begin{scope}[xshift=\offset]
        \ifnum\panel=0\relax
          \colorlet{edgeink}{black!85}
          \colorlet{vertexink}{black!85}
        \else
          \colorlet{edgeink}{black!35}
          \colorlet{vertexink}{black!45}
        \fi

        \coordinate (zC) at (2.81,2.01);
        \coordinate (zD) at (2.07,0.62);

        \coordinate (zCLabel) at (2.99,1.79);
        \coordinate (zDLabel) at (2.30,0.62);

        \path
          (0.00,1.36) coordinate (p)
          (0.16,2.55) coordinate (q)
          (1.07,1.60) coordinate (r)
          (1.11,2.81) coordinate (s)
          (1.74,3.81) coordinate (t)
          (2.66,3.19) coordinate (y)
          (2.89,4.29) coordinate (v)
          (4.14,2.00) coordinate (x)
          (4.59,3.82) coordinate (u)
          (5.29,2.84) coordinate (tip)
          (2.96,0.00) coordinate (tail)
          (3.97,0.44) coordinate (i)
          (5.21,1.40) coordinate (j)
          (5.88,0.52) coordinate (k)
          (6.41,1.70) coordinate (l);

        \path[cluster,draw=outlineC] (2.42,1.97)
          .. controls (2.37,1.30) and (3.32,1.09) .. (4.06,1.28)
          .. controls (4.52,1.38) and (4.44,1.89) .. (4.92,2.10)
          .. controls (5.41,2.29) and (5.82,2.32) .. (5.77,2.98)
          .. controls (5.72,3.76) and (5.10,4.36) .. (4.57,4.26)
          .. controls (3.95,4.19) and (4.21,3.50) .. (3.99,3.16)
          .. controls (3.72,2.74) and (2.42,2.78) .. cycle;

        \path[cluster,draw=outlineD] (1.69,.66)
          .. controls (1.71,1.20) and (2.24,1.12) .. (2.74,.82)
          .. controls (3.14,.59) and (3.42,.20) .. (3.34,-.23)
          .. controls (3.25,-.65) and (2.56,-.48) .. (2.09,-.08)
          .. controls (1.82,.15) and (1.66,.37) .. cycle;

        \draw[graph edge] (p)--(q)--(s)--(r)--cycle;
        \draw[graph edge] (q)--(r);
        \draw[graph edge] (p) .. controls (1.58,.94) and (2.02,2.17) .. (s);
        \draw[graph edge] (s)--(t)--(v)--(y)--cycle;
        \draw[graph edge] (t)--(y);
        \draw[graph edge] (i)--(j)--(l)--(k)--cycle;
        \draw[graph edge] (j)--(k);

        \draw[graph edge] (zC)--(x);
        \draw[graph edge] (x)--(u)--(tip)--cycle;
        \draw[graph edge] (zC)--(zD)--(x);
        \draw[graph edge] (zD)--(tail);

        \draw[graph edge] (s)--(zC)--(y)--(x);
        \draw[graph edge] (p)--(zD);
        \draw[graph edge] (j)--(x)--(i)--(tail);

        \ifnum\panel=0\relax
          \draw[graph edge] (u)--(v);
        \else
          \draw[black,line width=.95pt,dash pattern=on 3.3pt off 2.2pt]
            (v)--node[midway,above=2pt,vertex label] {$e$} (u);

          \draw[route,draw=toCenter]   (u)--(x)--(zC);
          \draw[route,draw=fromCenter] (zC)--(y)--(v);

          \path (u)--coordinate[midway] (firstArrow)  (x);
          \path (y)--coordinate[midway] (secondArrow) (v);
          \draw[->,>=stealth,toCenter,line width=1.6pt]   (u)--(firstArrow);
          \draw[->,>=stealth,fromCenter,line width=1.6pt] (y)--(secondArrow);
        \fi

        \foreach \name in {p,q,r,s,t,y,v,x,u,tip,tail,i,j,k,l}
          \node[vertex] at (\name) {};

        \ifnum\panel=1\relax
          \foreach \name in {u,x}
            \node[vertex,fill=toCenter] at (\name) {};
          \foreach \name in {y,v}
            \node[vertex,fill=fromCenter] at (\name) {};
          \node[vertex label,above right=2pt] at (u) {$u$};
          \node[vertex label,above left=2pt]  at (v) {$v$};
          \node[vertex label,right=5pt]       at (x) {$x$};
          \node[vertex label,below left=2pt]  at (y) {$y$};
        \fi

        \node[center vertex] at (zC) {};
        \node[center vertex] at (zD) {};
        \node[vertex label,anchor=west] at (2.99,1.79) {$z_C$};
        \node[vertex label,anchor=west] at (2.30,0.62) {$z_D$};
        \node[font=\large] at (5.61,4.13) {$C$};
        \node[font=\large] at (1.96,-.36) {$D$};

        \ifnum\panel=0\relax
          \node[anchor=base] at (3.20,-1.02)
            {{\small\textbf{(a)}\enspace Balanced cut $\mathcal C=\{C,D\}$}};
        \else
          \node[anchor=base] at (3.20,-1.02)
            {{\small\textbf{(b)}\enspace Routing $\rho_{\mathcal C}(e)$}};
        \fi
      \end{scope}
    }
  \end{tikzpicture}%
  }
  \caption{Hollow squares mark the cluster centers \(z_C\) and \(z_D\).
    In~(a), a balanced cut $\cC$ consisting of two clusters $C$ and $D$.  In~(b), the dashed cut edge \(e=(u,v)\)
    is rerouted along the blue path \(u\to x\to z_C\) in
    \(G[V(\mathcal C)]\), followed by the orange path \(z_C\to y\to v\)
    in \(G\).}
  \label{fig:balanced-cut-routing}
\end{figure}
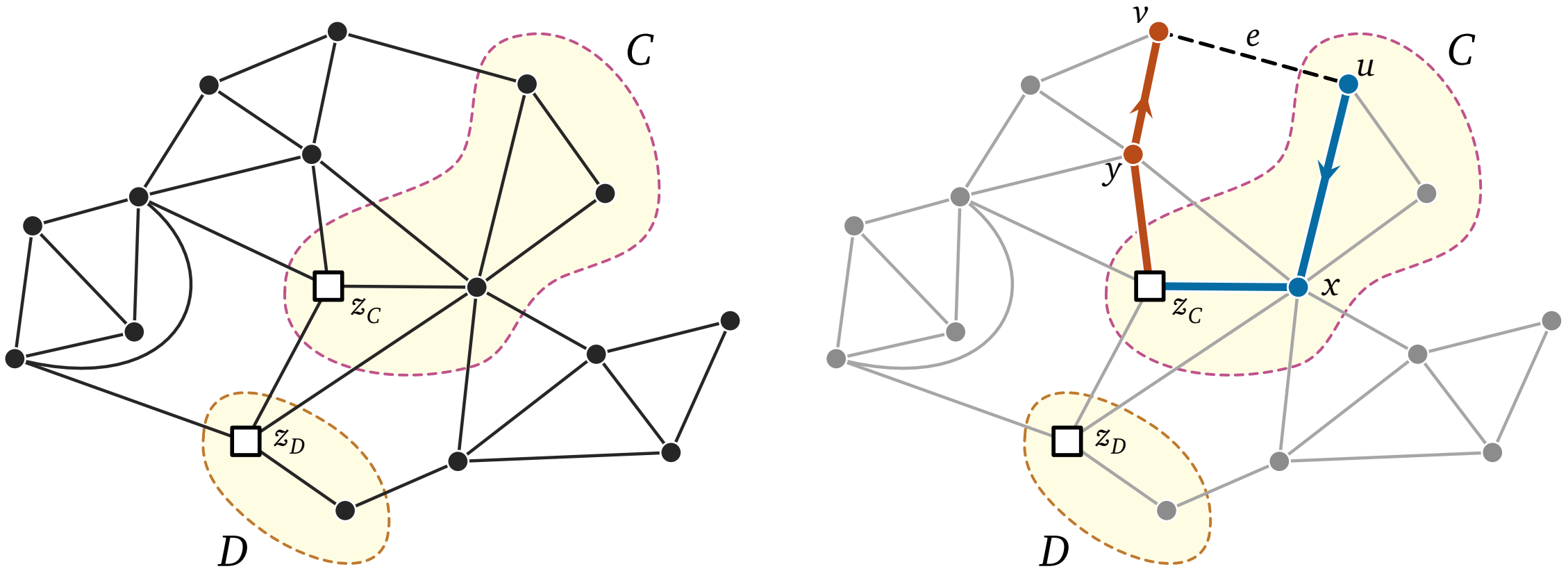

A subset of vertices $C$ is called a \EMPH{cluster} if the induced subgraph $G[C]$ is connected. For each cluster $C$ in this section, there is a single special vertex $z_C\in C$ called the \EMPH{center of $C$}. Let \EMPH{$\cC$} be a collection of vertex-disjoint clusters of $G$, and \EMPH{$Z_{\cC}$} be the set of centers. We denote by $V(\cC) = \cup_{C\in \cC}C$. For every edge $e = (u,v)$ such that exactly one of its endpoints belongs to $V(\cC)$, we say that \EMPH{$e$ is cut by $\cC$}. See \Cref{fig:balanced-cut-routing}(a).

\begin{definition}[Balanced Cut, Simplified]\label{def:balanced-cut-intro}  Let $G = (V,E)$ be an $n$-vertex planar graph. Let $\cC$ be a set of vertex-disjoint clusters. We say that $\cC$ is a \EMPH{balanced cut} of $G$ if  $|V(\cC)\setminus Z|\leq n/4$, and every connected component $O$ of $G\setminus \cC$ has $|V(O)| \leq n/4$. 
\end{definition}

We will be using balanced cuts in the treewidth embedding in a standard way~\cite{CCCLPP25}: construct a balanced cut $\cC$ of $G$, make a bag out of the centers of clusters in $\cC$, add edges from the centers to the endpoints of the edges cut by $\cC$, and recursively embed $G[V(\cC)]$ as well as connected components of $G\setminus \cC$. The balanced property of the cut ensures that the number of vertices decreases by a constant fraction at each step, so the recursion depth is $O(\log n)$. 

For every edge $e = (u,v)$ cut by $\cC$,  w.l.o.g., we assume that $u\in C$ for some cluster $C\in \cC$ and $v\not\in V(\cC)$. We must reroute  $e$ through the center $z_{C}$, inducing a \EMPH{routing scheme}, denoted by \EMPH{$\rho_{\cC}$}:
\begin{equation}
    \rho_{\cC}(e) \coloneqq \begin{cases}
  \pi_{G[V(\cC)]}(u,z_C)\circ \pi_{G}(z_C,v), & \text{$e = (u,v)$ is cut by $\cC$ and $u\in C$ for some cluster $C\in \cC$}, \\
    e,  & \text{otherwise }.
\end{cases} 
\end{equation}
where $\pi_G(x,y)$ denotes the shortest path from $x$ to $y$ in $G$. See \Cref{fig:balanced-cut-routing}(b).  The key technical contribution of this paper is the following lemma, which constructs a stochastic balanced cut with low expected routing cost.

\begin{restatable}[Stochastic Balanced Cuts, Simplified]{lemma}{simplifiedbalancedcuts}\label{lm:cuts-simplfiied} Let $G = (V,E,\ell)$ be any $n$-vertex planar graph with edge length function $\ell$. For any parameter $\eps \in (0,1)$, there is a distribution $\bbC$ of balanced cuts of $G$ such that (i) every cut $\cC$ in the support of $\bbC$ has $|\cC| = O(\log n/\eps)$ clusters and (ii) for every edge $e\in G$:
\begin{equation}
    \bbE_{\cC\sim \bbC}[\ell(\rho_{\cC}(e))] \leq (1+ \frac{\eps}{\log n})\ell(e)
\end{equation}
Furthermore, we can sample a cut from $\bbC$ in polynomial time. 
\end{restatable}

Essentially, \Cref{lm:cuts-simplfiied} says we can sample a balanced cut $\cC$ and route all edges cut by $\cC$ through its centers, paying only $(1+ \frac{\eps}{\log n})$. As we noted earlier, the embedding is recursive, and the recursion depth is $O(\log n)$. Thus, the accumulated total expected distortion for every edge $e\in G$ is:
\begin{equation}
    (1+ \frac{\eps}{\log n})^{O(\log n)}\ell(e) = (1+O(\eps))\ell(e)~,
\end{equation}
as desired. The final treewidth bound is $O(\log n/\eps)$ since $|\cC | = O(\log n/\eps)$. (In the actual construction, we have to be more careful in the recursion to avoid the treewidth accumulating over many levels.)

\paragraph{Previous balanced cuts.~} The stochastic balanced cut constructed in prior works~\cite{CLPP23,CCCLPP25} has the same distortion guarantee  but only gives $|\cC| = O(\log^3 n/\eps)$. (For simplicity, we assume that the aspect ratio $\Phi = n^{O(1)}$.)  This is due to the technique they used to construct such a cut. First, they construct a hierarchy \EMPH{$\frakH$}  of clusters where clusters at level $i$ have diameter roughly $2^{i}$. The hierarchy is stochastic so that every edge $(u,v)$ is cut by a level-$i$ cluster with probability $O(d_G(u,v)/2^i)$.  Then they constructed a distribution $\Tilde{\bbC}$ of cuts such that every cluster in \EMPH{$\frakH$} belongs to  $\Tilde{\bbC}$ with probability at most $ 1/\psi \coloneqq \eps/(\log^2 n)$. The $\log^2(n)$ factor is because now an edge could be cut in $O(\log^2 n)$ events: it was potentially cut by $\log(n)$ (inner) levels of $\frakH$, and another $\log(n)$ (outer) levels of the recursion in the embedding. Taking all of these events into account, the expected detour cost is $O(\log^2n)(\frac{1}{\psi})2^i\cdot O(d_G(u,v)/2^i) = O(\eps)d_G(u,v)$, implying expected distortion $(1+O(\eps))d_G(u,v)$.   Then they show that the size of the balanced cut $\cC$ is at most $\tau = O(\log n \cdot \psi) = O(\log^3 n/\eps)$. (We use parameters $\tau$ and $\psi$ as exactly defined in~\cite{CCCLPP25}.) Thus, compared to the optimal bound $\log(n)/\eps$, one extra log factor comes from the fact that the clusters are constructed from the hierarchy $\frakH$ with depth $\log(n)$, and another factor comes from the relationship between $\tau$ and $\psi$; both are tight~\cite{CCCLPP25}. Furthermore, since the construction uses a clustering hierarchy of depth $O(\log(\Phi))$, the dependency on the aspect ratio naturally kicks in the final treewidth bound.

\paragraph{Our balanced cuts: the overview of the proof.~} Our construction in \Cref{lm:cuts-simplfiied} abandons the clustering hierarchy $\frakH$ entirely. We no longer construct clusters with diameter $2^i$ such that an edge  $(u,v)$ is cut by one of the clusters with probability $O(d_G(u,v)/2^i)$. Instead, we use Yao's minimax principle: for any non-negative weight function $w$ on the edges of $G$, we construct a balanced cut with small total weighted detour. This construction significantly deviates from what is currently used in all known embeddings of graphs and metric spaces into low-treewidth graphs~\cite{Tal04,FFKP18,CLPP23,CCCLPP25}.

\begin{lemma}\label{lm:minimax-weight-simplified} Let $G = (V,E,\ell)$ be an $n$-vertex planar graph and $\eps\in(0,1)$ be a parameter. For any edge weight function $w: E\rightarrow \real_{\geq 0}$, one can construct in polynomial time a balanced cut $\cC$ such that:
\begin{equation*}
    \sum_{e\in E} w(e) \left(\ell(\rho_{\cC}(e)) - \ell(e)\right) \leq \frac{\eps}{\log n}\sum_{e\in E} w(e)\ell(e)
\end{equation*}
and $|\cC| = O(\log(n)/\eps)$.
\end{lemma}

By Yao's minimax principle, \Cref{lm:minimax-weight-simplified} implies the existence of a stochastic balanced cut in \Cref{lm:cuts-simplfiied}. The only catch is that this principle does not give a sampling algorithm. For this, we apply the technique of Filtser and Le~\cite{FL21} to develop an algorithmic minimax (\Cref{subsec:minimax}) and get a polynomial-time sampling scheme.

Now we sketch the proof of \Cref{lm:minimax-weight-simplified}. Let \EMPH{$k = \Theta(\log n/\eps)$}, and \EMPH{$\mu(E) = \sum_{e\in E} w(e)\ell(e)$}. For every subset  $X$ of edges of $E$, we define $\mu(X) = \sum_{e\in X} w(e)\ell(e)$.  A key idea in the proof is to define for every vertex $v$ a radius $\rho(v)$ such that the total $\mu$-measure of the edges inside the ball $B(v,\rho(v))$ is
\begin{equation}\label{eq:total-path-measure}
    \mu(B(v,\rho(v))) \approx \mu(E)/k^2 = O((\eps/\log n)^2)\mu(E)
\end{equation}
On a more technical note, the definition of edges inside the ball is somewhat delicate\footnote{More precisely, we view edges as continuous, and a ball could cut an edge into two segments, with one segment considered inside the ball.}, but any edge inside the ball has both endpoints in the ball. 

By definition, there are only $O(k^2)$ disjoint balls in $\{B(v,\rho(v))\}_{v\in V}$ since their total $\mu$-measure adds up to at most $\mu(E)$.  By a result of Bousquet, Cames van Batenburg, Esperet, Joret, Lochet, Muller, and Pirot~\cite{BCEJLMP21}, we can find a set \EMPH{$Z^+\subseteq V$ of size $O(k^2)$} such that:
\begin{equation}
 Z^+\cap B(v,\rho(v)) \not=\varnothing \qquad \forall v\in V   
\end{equation}

Let $F_{Z^+}$ be the shortest-path spanning forest rooted at $Z^+$, where ties are broken consistently. By a variant of the planar separator theorem (\Cref{lm:forest-separator}), we can construct a balanced separator $\cP$ of $O(\sqrt{|Z^+|}) = O(k)$ paths rooted at $O(k)$ vertices in $Z^+$. Let $Z$ be the roots of these paths, which would then be the centers of the clusters in our final balanced cut. The key point here is that, the total $\mu$-measure all of the union of the $k$ paths in $\cP$ is at most:
\begin{equation}
    O(k \cdot \mu(E)/k^2) = O(\mu(E)/k) = O(\eps/\log n) \sum_{e\in E}w(e)\ell(e)
\end{equation}
This is an important starting point: if we reroute cut edges along paths in $\cP$, and each path is used only a few times, then the total routing cost is already within check. (Of course, there is no reason to expect that each path is only used a few times.)  

The next key idea is to define a function $\phi(v)$, which is a normalized distance function from $v$ to paths in $\cP$. Specifically, let $P^{\cup} = \cup_{Q\in \cP}Q$, and:

\begin{equation}\label{eq:dist}
 \phi(v) =  \begin{cases}
    d_{G}(v,P^{\cup})/d_{G}(v,Z^+) & \text{if } d_{G}(v,Z^+)\not= 0  \\
    +\infty      & \text{otherwise}
\end{cases} 
\end{equation}
Basically, $\phi(x)$ is the distance from $v$ to $P^{\cup}$, normalized by the distance from $v$  to $Z^+$. This function behaves like a distance function (in an important regime): for any vertex $u$ on the shortest path from $v$ to $P^{\cup}$, $\phi(u) < \phi(v)$. The two most important properties of this function that we will use in our proof are:
\begin{enumerate}
    \item Let $B_{\phi}(\cP,1/4)$ be ball of radius $1/4$ from the rooted path separator $\cP$ w.r.t distance function $\phi$. Then $\mu(B_{\phi}(\cP,1/4) = O(\frac{1}{k})\mu(E)$. (\Cref{lm:mu-U} in the proof.) That is, in the worst case, the total measure of $B_{\phi}(\cP,1/4)$ is within a constant factor of the total measure of the paths in $\cP$ in \Cref{eq:total-path-measure}.
    \item For any radius threshold $\tau\in [1/8,1/4]$, vertices in $B_{\phi}(\cP,1/4)$ induce a balanced cut \EMPH{$\cC_{\tau}$} with $O(k) = O(\log n/\eps)$ clusters, one cluster for each path in $\cP$. (\Cref{lm:sept-tau} in the proof.)
 \end{enumerate}

 Using the two properties, we can show (via elementary calculation) that by choosing a random $\tau \in [1/8/1,4]$, the expected total routing cost along $\rho_{\tau}$, the routing scheme induced by $\cC_{\tau}$, is:
    \begin{equation}
        \bbE_{\tau}[\sum_{e}w(e)(\rho_{\tau}(e) - e ))] \leq  \frac{O(1)}{k} \sum_{e}w(e)\ell(e) = \frac{\eps}{\log n} \sum_{e}w(e)\ell(e).
    \end{equation}
 This means we can find a threshold $\tau \in [1/8/1,4]$ where the total weighted detour cost is at most $\frac{\eps}{\log n} \sum_{e}w(e)\ell(e)$,  giving a cut $\cC_{\tau}$ satisfying \Cref{lm:minimax-weight-simplified}.

Finally, we remark that there are two subtleties that we sweep under the rug in this overview: (1) the construction of the balanced cuts has to take into account the boundary vertices passed down from ancestral recursive calls and (2) we incorporate vertex measure, not just edge measure $\mu$, in the definition of $\rho(v)$, to guarantee that the cut is balanced. These technical details do not change the overall idea of the proof.

\paragraph{Extension to minor-free graphs.} For minor-free graphs, we extend \Cref{lm:minimax-weight-simplified} using a variant of the Robertson--Seymour decomposition~\cite{CLPP23}. By algorithmic minimax, we obtain a stochastic balanced cut as in \Cref{lm:cuts-simplfiied} for minor-free graphs. Given the cut, our algorithm only uses the minor-closed property and hence extends without changes to minor-free graphs.  

\section{Preliminaries}

Let $G = (V,E,\ell)$ be a connected undirected graph with vertex set $V$, edge set $E$, and non-negative edge length function $\ell: E \rightarrow\real_{\geq 0}$. Let $d_G$ be the shortest path distance function.  Graphs in this paper are connected, unless explicitly stated otherwise. If the vertex set, edge set, and length function are not explicitly defined, we denote them by $V(G)$, $E(G)$, and $\ell_G$, respectively.  Sometimes we also assign weights to vertices and edges of $G$, and in this case, we use $\omega$ to denote vertex weights and $w$ to denote edge weights.  The shortest distances are distances w.r.t. edge length $\ell$.

A stochastic embedding of a weighted graph $G$ is a distribution $\cD$ over weighted graphs $H$ and maps $f:V(G)\to V(H)$. The embedding is \EMPH{dominating} if  for every graph $H$ in the support of $\cD$ and every two vertices $u,v\in V(G)$, $d_G(u,v)\leq d_H(u,v)$. Embeddings in this paper are dominating.  We say that $\cD$ has expected distortion $t$ if:
\begin{equation}
    \mathbb{E}\!\left[d_H(f(u),f(v))\right]\leq t\cdot d_G(u,v).
\end{equation}

For a vertex $x\in V$, let $N_G(x)$ be the set of neighbors of $x$ in $G$, and $N_G[x] = N_G(x)\cup \{x\}$.  For a subset of vertices $X\subseteq V$, we define the closed neighborhood to be $N_G[X] = \cup_{x\in X}N_G[x]$ and the open neighborhood to be $N_G(X) = N_G[X]\setminus X$. If the graph is clear from context, we drop the subscript $G$ from the neighborhood notation.

Fix a shortest path $\pi_G(u,v)$ for every two vertices $u$ to $v$ in $G$. For a path $P$ and two vertices $u,v\in P$, we denote by $P[u,v]$ the subpath from $u$ to $v$.  For any two paths $P[u,v]$ and $Q[v,z]$, we denote by $P[u,v]\circ Q[v,z]$ the walk from $u$ to $z$ obtained by concatenating $P[u,v]$ and $Q[v,z]$ at $v$.

\paragraph{Tree decomposition.}  We say that a tree $\cT$ is a tree decomposition of $G = (V,E)$ if nodes of $\cT$ are associated with subsets of $V$, called \EMPH{bags}, such that:
\begin{enumerate}
    \item Every vertex $v\in V$ belongs to some bag of $\cT$.
    \item  Every edge $(u,v)\in E$, some bag contains both $u$ and $v$.
    \item  For every vertex $v\in V$, all the bags containing $v$ induce a connected subtree of $\cT$.
\end{enumerate}

The width of $\cT = \max_{X\in \cV(\cT)}|X|-1$, where $\cV(\cT)$ is the set of bags in $\cT$.  The treewidth of $G$ is the minimum width over all tree decompositions of $G$.

We will use two elementary known results to construct stochastic balanced cuts in planar and minor-free graphs: rooted path separators of planar graphs and small ball hitting sets.

\paragraph{Rooted path separators in planar graphs.}  This new variant of the separator theorem basically says that given a spanning forest $F$ rooted at $s$ vertices, we can construct a balanced separator containing $O(\sqrt{s})$  leaf-to-root paths from $F$.

\begin{restatable}{lemma}{forestseparator}\label{lm:forest-separator-prelim}
Let $G = (V,E)$ be a planar graph with a non-negative vertex weight function $\omega$, and $F$ be a rooted spanning forest that has $s$ connected components. Then there exists a balanced separator $\cP$ of $G$ w.r.t.\ $\omega$ consisting of $O(\sqrt{s})$ rooted paths of $F$ such that for every connected component $C\in G\setminus \cP$,
\[
    \sum_{v\in C}\omega(v) \leq \omega(V)/2,
\]
where $\omega(V) = \sum_{v\in V}\omega(v)$.
\end{restatable}

Though we are not aware of the exact form of \Cref{lm:forest-separator} in published papers, we can easily derive it from known results in the literature, for example, the separator theorem by Cai, Har-Peled, and Ye~\cite{CHY18}; we give the details of the proof in \Cref{sec:apps}.

\paragraph{Small hitting sets for balls in minor-free graphs.~} Bousquet, Cames van Batenburg, Esperet, Joret, Lochet, Muller, and Pirot showed that 
balls in minor-free graphs have small hitting sets: the size of the hitting set is proportional to the number of vertex-disjoint balls. Their theorem is stated only for unweighted graphs, but it holds for graphs with edge lengths as well.

\begin{theorem}[Theorem 2 in~\cite{BCEJLMP21}, adapted]\label{thm:ball-traversal} Let $\cB$ be a finite collection of balls in a $K_h$-minor-free graphs graph $G = (V,E,\ell)$ with edge lengths. Let $\cI\subseteq \cB$ be a maximum cardinality set of vertex-disjoint balls in $\cB$. There exists a \EMPH{$\cB$-hitting set} $Z\subseteq V$ such that $|Z| = O_h(|\cI|)$ and for every ball $B\in \cB$, $Z\cap B\not=\varnothing$.
\end{theorem}

We note that $Z$ depends on $h$ exponentially. 

\section{The Embedding Algorithm}

A key technical concept in the embedding algorithm is the stochastic balanced cut with a small routing cost. Let \EMPH{$\cC$} be a collection of vertex-disjoint clusters of $G$, and \EMPH{$Z_{\cC} = \{z_C: C\in \cC\}$} be the set of centers, one center for each cluster. We denote by $V(\cC) = \cup_{C\in \cC}C$. We now define a $B$-balanced cut where $B$ is a subset of vertices called boundary vertices. This definition is slightly more general than the definition of a balanced cut in \Cref{subsec:tech-ideas}.

\begin{definition}[Balanced Cut]\label{def:balanced-cut}  Let $G = (V,E)$ be an $n$-vertex graph. Let $B\subseteq V$ be a designated subset of vertices, called the \EMPH{boundary vertices} of $G$. Let $\cC$ be a set of vertex-disjoint clusters, each cluster $C\in \cC$ with a center vertex $z_C$, and $Z = \{z_{C}\}_{z\in C}$ be the set of centers. We say that $\cC$ is a \EMPH{balanced $B$-cut} of $G$ if:

\begin{enumerate} 
    \item \textbf{($B$-nearly-disjoint)~} For every cluster $C\in \cC$, the only vertex of $C$ that can belong to $B$ is its center. 
    \item  \textbf{($B$-balanced and $V$-balanced).~}  $|V(\cC)\setminus Z|\leq n/4$, and every connected component $O$ of $G\setminus \cC$ has
     \begin{equation}
        |V(O)| \leq n/4, \qquad \text{and}\qquad |B\cap V(O)| \leq |B|/4
    \end{equation}
\end{enumerate}
\end{definition}

We will be using balanced cuts in the treewidth embedding in a standard way~\cite{CLPP23,CCCLPP25}: construct a balanced cut $\cC$ of $G$, make a bag out of the center set $R_{\cC}$ of clusters in $\cC$, and recursively embed $G[V(\cC)]$ as well as (graphs constructed from) connected components of $G\setminus \cC$. The balanced property of the cut ensures that the number of vertices decreases by a constant fraction at each step, so the recursion depth is $O(\log n)$. However, for every edge $e = (u,v)$ such that one of its endpoints, say $u$, belongs to a cluster $C\in \cC$, while the other endpoint $v\not\in V(\cC)$,  we must reroute it through the center $z_{C}$. We say that such an edge \EMPH{$e$ is cut by $\cC$}.  The rerouting of the edges cut by $\cC$ gives rise to a \EMPH{routing scheme induced by $\cC$}, denoted by \EMPH{$\rho_{\cC}$}, defined as follows:
\begin{equation}\label{eq:rho}
    \rho_{\cC}(e) \coloneqq \begin{cases}
  \pi_{G[V(\cC)]}(u,z_C)\circ \pi_{G}(z_C,v), & \text{$e = (u,v)$ is cut by $\cC$ and $u\in C$ for some cluster $C\in \cC$}, \\
    e,  & \text{otherwise }.
\end{cases} 
\end{equation}
Note that  $\pi_{G}(z_C,v)$ in the definition of  $\rho_{\cC}(e)$ could contain a vertex in $V(\cC)$, while intuitively we would want this path to be contained entirely in $G[Z_{\cC}\cup O]$ for the recursion to handle it, where $O$ is the connected component of $G\setminus \cC$ containing $v$. (Indeed, the prefix $ \pi_{G[V(\cC)]}(u,z_C)$ is handled by the recursion on $G[V(\cC)]$.)  In our embedding algorithm, we resolve this issue by adding a direct edge from $z_C$ to $v$ in the host graph $H$ with length exactly $d_G(z_C,v)$. Thus, the fact that  $\pi_{G}(z_C,v)$ is not contained entirely in  $G[Z_{\cC}\cup B]$ is not a problem.

For a low-distortion embedding, we want $\rho_{\cC}(e)$ to be not much larger than $\ell(e)$ in expectation. A key technical contribution of this paper is to design such a distribution of balanced cuts, formalized in the following lemma.

\begin{restatable}[Stochastic Balanced Cuts with Low Distortion]{lemma}{stochasticbalancedcuts}\label{lm:distribution-cuts}
Let $k$ be any given parameter sufficiently larger than $1$. Let $G = (V,E,\ell)$ be any $n$-vertex planar graph with edge length function $\ell$. $B\subseteq V$ be a designated set of boundary vertices such that $|B|\leq k^2$. Then there is a distribution $\bbC$ of balanced $B$-cuts of $G$ such that (i) every cut $\cC$ in the support of $\bbC$ has $|\cC| = O(k)$ clusters and (ii) for every edge $e\in G$:
\begin{equation}\label{eq:expected-dis}
    \bbE_{\cC\sim \bbC}[\ell(\rho_{\cC}(e))] \leq (1+ \frac{O(1)}{k})\ell(e)
\end{equation}
Furthermore, the support of $\bbC$ has only $O(n\log n)$ balanced cuts, which can be constructed in polynomial time, and sampling is uniform at random in $O(1)$ time. 
\end{restatable}

The condition $B\leq k^2$ mainly arises because we want clusters in a balanced cut to avoid $B$ (except for their centers). In our embedding, $B = O(k)$, so $B\leq k^2$ holds automatically as long as $k$ is sufficiently large.

Specifically, we set $k = \frac{\eta\log n}{\eps}$ in \Cref{lm:distribution-cuts} for a sufficiently large constant $\eta$, so the expected distortion (at one level of the recursion) is about $(1+\frac{\eps}{4\log n})$. Over $\ceil{\log(n)}$ levels, the expected distortion is $(1 + \frac{\eps}{4\log n})^{\ceil{\log(n)}} \leq 1+\eps$. The cut $\cC$ sampled from $\bbC$ will has  $|Z_{\cC}| \leq O(k) = O(\log n/\eps)$ centers.  Thus, the bag from $Z_{\cC}$ has size $O(\log n/\eps)$, which is also our final treewidth bound. The fact that $\cC$ is $B$-balanced (item 2 of \Cref{def:balanced-cut}) allows us to control the size of $B$ to be always of size $ O(k) = O(\log n /\eps)$ so that $B\leq k^2$ holds when $\eta$ is sufficiently large. 

Next, we describe the embedding algorithm, given the stochastic balanced cuts in \Cref{lm:distribution-cuts}. The algorithm is similar to those in prior work, using their own version of stochastic balanced cuts~\cite{CLPP23,CCCLPP25}. Readers who are already familiar with previous work are encouraged to skip to \Cref{sec:balanced-cuts} where we prove \Cref{lm:distribution-cuts}.

\paragraph{The embedding algorithm.~} We assume that the length of every edge of $G$ is the shortest distance between its endpoints; this can be done by reassigning the edge length to be the endpoints' distance.  The pseudo-code of the recursive embedding algorithm is in \Cref{alg:embed}. Procedure $\textsc{Embed}(G,B)$ takes a planar graph $G = (V,E)$ and its boundary vertex set $B$ as input, and produces a graph $H$ with $V(H) = V$ and a tree decomposition $\cT$ of $H$. The root bag of $\cT$ will contain $B$.   Initially, for the input graph, $B = \varnothing$.  We choose a sufficiently large constant $\eta$ to be used in all recursive calls. The target treewidth bound is $O(\log n/\eps)$.  The base case is when the number of vertices is small, namely, $n\leq 8\eta^2 \log(n)/\eps$: we simply make $H$ a complete graph with all pairwise edges and set $\cT$ to be a single-bag tree decomposition.

For the non-base case, we sample a $B$-balanced cut $\cC$ by applying \Cref{lm:distribution-cuts} with $k = \eta\log(n)/\eps$. We initialize the tree decomposition $\cT$ with a single bag $B\cup Z_{\cC}$, where $Z_{\cC}$ is the set of centers of $\cC$. Then we recursively embed connected components of $G[V(\cC)]$ and $G\setminus \cC$. Specifically:
\begin{itemize}
    \item \textbf{Type-I recursion.~}  For each connected component $A\in G[V(\cC)]$ in the balanced cut, $A$ is a union of clusters since each cluster is connected. The boundary vertices of $A$ are the centers of the clusters in $A$. We note that only the boundary vertices of $A$ could contain vertices in $B$ by item 1 of \Cref{def:balanced-cut}. We recursively call \textsc{Embed}$(A,Z_{\cC}\cap A)$ to get an embedding of $A$ into a graph $H_A$ and its tree decomposition $\cT_A$. Then we attach $\cT_A$ to $\cT$ by making the root of $\cT_A$ a child bag of $B\cup Z_{\cC}$, the root bag of $\cT$, and add $H_A$ to $H$. Observe that:
    \begin{equation}\label{eq:obs-rootA}
        \text{(root-bag-of-$\cT_A$)}\cap \text{(root-bag-of-$\cT$)} = Z_{\cC}\cap A
    \end{equation}
    \item   \textbf{Type-II recursion.~}  For each connected component $O\in G\setminus \cC$, we want to recursively embed not just edges in $O$. To see this, consider an edge $e = (u,v)$ cut by $\cC$, i.e., $u\in C$ for some cluster $C\in \cC$ and $v\in O$. The rerouted path $\rho_{\cC}(e)$ goes through the center $z_C\in C$. The prefix subpath from $u$ to $z_C$ will be approximately preserved (in expectation) by the recursive calls on connected components of $G[V(\cC)]$ above. However, the suffix path from $z_C$ to $v$ cannot be preserved simply by recursing on $O$, since $O$ does not contain $z_C$. 
    
    First, we contract every cluster into its center to obtain a graph $G^{-}$, and remove parallel edges and loops.  Note that $G^{-}$ is a planar graph.  Then for every edge  $(z_C,v)\in G^{-}$ that is not in $G$, we set the length \EMPH{$\ell(z_C,v) = d_G(z_C,v)$ in $G^{-}$}. Next, for each connected component $O\in G\setminus \cC$, let $Z_O$ be the set of centers in $G^-$ that have neighbors in $O$. We recursively call \textsc{Embed}$(G^{-}[V(O)\cup B_O],B_O)$  to embed $G^{-}[V(O)\cup B_O]$ with the boundary vertex set \EMPH{$B_O = Z_O\cup (V(O)\cap B)$}. Here, $B_O$ contains all boundary vertices of $G$ in $O$ and vertices in $Z_O$. Let $H_O$ be the output host graph and $\cT_O$ be the tree decomposition of $H_O$ from the recursion. We simply add $H_O$ to $H$ and attach $\cT_O$ to  $T$ by making the root bag of $\cT_O$ a child bag of the root bag of $\cT$. We have:

    \begin{equation}\label{eq:obs-rootO}
        \text{(root-bag-of-$\cT_O$)}\cap \text{(root-bag-of-$\cT$)} = B_O
    \end{equation}

    Back to the discussion on the cut edge $e= (u,v)$, the suffix path $\pi_G(z_C,v)$ now becomes an edge $(z_C,v)$ in $V(O)\cup B_O$ with length $\ell(\pi_G(z_C,v)) = d_G(z_C,v)$. The edge will be approximately preserved by the recursion. As both subpaths of $\rho_{\cC}(e)$ have low distortion, by \Cref{lm:distribution-cuts}, we can show that $e$ has a small expected distortion.
\end{itemize}
We note that every edge $(u,v)\in H$ is assigned a length to be $d_G(u,v)$. 

\begin{algorithm}[H]
\caption{$\textsc{Embed}(G,B)$}
\label{alg:embed}
\textbf{Input:} A planar graph $G = (V,E)$, and $B\subseteq V$ the set of boundary vertices. The constant $\eta$ is sufficiently large, and $n$ is the number of vertices of $G$.\\
\textbf{Output:} A tree decomposition $\cT$ of $G$ whose root bag contains $B$.
\begin{algorithmic}[1]
    \If{$n\leq 8\eta^2 \log(n)/\eps$} \Comment{\lightgray{the base case}}
        \State $H\leftarrow (V, {V\choose 2})$
        \State $\cT\leftarrow \{V\}$, a single-bag tree decomposition.
        \State return $(H,\cT)$
    \EndIf
    \State $\cC\overset{\text{sample}}{\leftarrow}\bbC$ where $\bbC$ is a distribution of $(B,\eta \log(n)/\eps)$-balanced cuts in \Cref{lm:distribution-cuts}. 
    \State $Z_{\cC}\leftarrow$ centers of clusters in $\cC$
    \State $H\leftarrow (V, {B\cup Z_{\cC}\choose 2})$ \Comment{\lightgray{$H$ contains edges between $B\cup Z_{\cC}$}} 
    \State $\cT\leftarrow \{B\cup  Z_{\cC}\}$\label{line:init-T} \For{ each connected component $A\in G[V(\cC)]$}
        \State $(\cT_A,H_A)\leftarrow$ \textsc{Embed}$(A,Z_{\cC}\cap A)$ \label{line:cTA}\Comment{\lightgray{The root of $\cT_A$ is a bag $Z_{\cC}\cap A$}}
        \State make the root of $\cT_A$ a child of $B\cup  Z_{\cC}$, effectively attaching $\cT_A$ to $\cT$\label{line:attaching-I}
        \State $H\leftarrow H\cup H_A$ 
    \EndFor
    \State $G^-\leftarrow$ graph obtained from $G$ by contracting each cluster $C$ in $\cC$ into $z_C$ \label{line:contract} \Comment{\lightgray{$z_C$ is the center of $C$}}
    \For{each connected component $O\in G\setminus \cC$}
        \State $Z_O\leftarrow N_{G^-}[O]\cap Z_{\cC}$ \Comment{\lightgray{$Z_O$ contains all centers adjacent to $O$ in $G^{-}$}}
        \State $B_O \leftarrow Z_O\cup (V(O)\cap B)$
        \State $(\cT_O,H_O) \leftarrow$ \textsc{Embed}$(G^{-}[V(O)\cup B_O],B_O)$ \label{line:cT-o}
        \State make the root bag of $\cT_O$ a child of the bag $B\cup  Z_{\cC}$, effectively attaching $\cT_O$ to $\cT$ \label{line:attaching-II}
       \State $H\leftarrow H\cup H_O$
    \EndFor
    \State return $(H,\cT)$
\end{algorithmic}
\end{algorithm}

First we show that the boundary set $B$ will never get too large.

\begin{lemma}\label{lm:B-size} Suppose that we call \Cref{alg:embed} on the input graph with $B = \varnothing$. Then in any recursive call:
\begin{equation}\label{eq:B-size}
    |B| \leq 2\eta^2\log(n)/\eps
\end{equation}
for a sufficiently large constant $\eta$.
\end{lemma}
\begin{proof}
We prove by induction.   The base case is at the top level of the recursion, where $B = \varnothing$. Since $\cC$ is a balanced $(B,\eta  \log(n)/\eps)$-cut, by \Cref{def:balanced-cut}, we have:
\begin{equation}\label{eq:cut-size}
    |Z_{\cC}| = |\cC| = O(k) =  O(\eta  \log(n)/\eps)\leq \eta^2  \log(n)/\eps
\end{equation}
for a sufficiently large constant $\eta$.  For Type-I recursion, the set of boundary vertices is a subset of $Z_{\cC}$ and hence has size at most $\eta^2  \log(n)/\eps < 2\eta^2\log(n)/\eps$. For Type-II recursion, we have $|Z_O|\leq |Z_{\cC}|$ while by \Cref{def:balanced-cut}, $|V(O)\cap B| \leq B/4$, which gives:
\begin{equation*}
    |B_O| \leq |Z_O| + |V(O)\cap B| \leq \eta^2\log(n)/\eps + |B|/4 \leq 2\eta^2\log(n)/\eps~,
\end{equation*}
which completes the proof.
\end{proof}

\mainthm*
\begin{proof} The embedding is $(H,\cT) = \textsc{Embed}(G,\varnothing)$.  For the analysis, we first bound the number of recursion levels, denoted by $L$. Then we bound the expected distortion and the treewidth by induction on the number of levels. 

 If we are in a non-base case of the recursion, $n\geq 8\eta^2 \log(n)/\eps$.  Observe that every time we recurse, the size of the graph is at most:
\begin{equation}
    n/4 + \text{(number of boundary vertices)} \leq n +  2\eta^2\log(n)/\eps \leq n/4 + n/4  = n/2
\end{equation}

For Type-I, we count the noncenters in the union and then the centers. For Type-II, we count $V(O)$ and then $Z_O$. Thus, $L \leq \ceil{\log_2n}$.

We observe that $V(H) = V(G)$, and every time we add an edge $(u,v)$ to $H$, its length in $H$ is the shortest distance between its endpoints in $G$. Thus, by the triangle inequality, the embedding is dominating. 

Now we bound the expected distortion by induction. We note that the embedding is dominating since $V(H) = V(G)$, and every time we add an edge $(u,v)$ to $H$, its length in $H$ is the shortest distance between its endpoints in $G$. 

We say that a recursive call is \EMPH{at level $i$} if its distance to the root of the recursion tree is exactly $i$. We prove that if the call $\textsc{Embed}(G,B)$ is at level $i$, then for every two vertices $u,v\in G$:
\begin{equation}\label{eq:distortion-induction}
   \bbE[d_H(u,v)]\leq (1 + \frac{\eps}{4\log n})^{L-i}d_G(u,v)
\end{equation}
The base case is $i = L$; in this case, $H$ is a complete graph, so $d_H(u,v) = d_G(u,v)$, and \Cref{eq:distortion-induction} holds.  

For the non-base case, by the triangle inequality, it suffices to show \Cref{eq:distortion-induction} when $e = (u,v)$ is an edge of $G$, and in this case, $d_G(u,v) = \ell(u,v)$.  Observe that if $e$ is not cut by $\cC$, then  $e$ is passed down to one of the recursive calls and hence \Cref{eq:distortion-induction} follows from the induction hypothesis. Otherwise, $e$ is cut by $\cC$. By the definition of being cut, $e = (u,v)$ where $u \in C$ for some cluster $C\in \cC$ and $v\in O$ for some connected component $O\in G\setminus \cC$.  Observe that the prefix of $\rho_{\cC}(e)$, namely $\pi_[G[V(\cC)]](u,z_C)$ is passed down to a connected component $A\in G[V(\cC)]$ in the Type-I recursion. By the induction hypothesis, we have:

\begin{equation}\label{eq:prefix-dist}
    \bbE[d_{H_A}(u,z_C)| \cC] \leq  (1+\frac{\eps}{4\log n})^{{L-(i+1)}}d_A(u,z_C)
\end{equation}
The suffix of $\rho_{\cC}(e)$, namely $\pi_G(z_C,v)$ is replaced by an edge $(z_C,v)$ in $G^{-}$ and hence is passed down to a Type-II recursive call on some connected component $O$ of $G\setminus \cC$, giving:
\begin{equation}\label{eq:suffix-dist}
    \bbE[d_{H_O}(z_C,v)  | \cC] \leq  (1+\frac{\eps}{4\log n})^{{L-(i+1)}}d_{H_O}(z_C,v)
\end{equation}

Combining \Cref{eq:prefix-dist} and \Cref{eq:suffix-dist}, we obtain:
\begin{equation}
    \bbE[\ell(\rho_{\cC}(e))| \cC] \leq  (1+\frac{\eps}{4\log n})^{{L-(i+1)}}\rho_{\cC}(e)
\end{equation}
and hence:
\begin{equation}
  \begin{split}
       \bbE[\ell(\rho_{\cC}(e))] &=  \bbE\left[\bbE[\ell(\rho_{\cC}(e))| \cC]\right] \\ 
       &\leq  (1+\frac{\eps}{4\log n})^{{L-(i+1)}}\bbE[\rho_{\cC}(e)] \\
       &\leq  (1+\frac{\eps}{4\log n})^{{L-(i+1)}}(1+ \frac{\eps}{4\log(n)}) \ell(e) \qquad  \text{(by \Cref{lm:distribution-cuts} and choosing $\eta$ sufficiently large)}\\
       & = (1+\frac{\eps}{4\log n})^{{L-i}}\ell(e)~,
  \end{split}
\end{equation}
giving \Cref{eq:distortion-induction}.

At the top level of the recursion $i = 0$, and hence the expected distortion is:
\begin{equation}
    (1+\frac{\eps}{4\log n})^L \leq e^{\eps L/4\log n} \leq e^{\eps/2} \leq (1+\eps)
\end{equation}
since $\eps \in (0,1)$.

Finally, we show that $\cT$ is a valid tree decomposition of $H$ and bound its treewidth. Note that every time we add an edge to $H$, both of its endpoints are in the same bag of $\cT$. Thus, it remains to show that for every vertex $v\in V(H)$, the bags containing $v$ induce a connected subtree of $\cT$, and we do so by induction. If $v\not\in Z_{\cC}\cup B$, then $v$ is contained in exactly one graph among the graphs in the recursive calls and hence, by induction, the bags containing $v$ induce a connected subtree of the tree decomposition returned from the recursive call. Otherwise,  $v\in Z_{\cC}\cup B$. Then $v$ is also a boundary vertex of the graphs in the recursive calls. By \Cref{eq:obs-rootA} and \Cref{eq:obs-rootO}, the returned tree decompositions have root bags that contain $v$ if $v$ is passed down to the recursive calls. Thus, by making the root of the returned tree decomposition a child bag of $B\cup Z_{\cC}$ in line~\ref{line:attaching-I} and line~\ref{line:attaching-II}, the bags containing $v$ of $\cT$ also induce a connected subtree.

Now we show by induction that the treewidth of $\cT$ is at most $8\eta^2 \log(n)/\eps = O(\log n/\eps)$. This holds trivially for the base case when $n\leq 8\eta^2 \log(n)/\eps$. For the non-base case, observe that $\cT$ is initialized in line~\ref{line:init-T} as a single bag of size at most:
\begin{equation}
    |B|  + |Z_{\cC}|\leq \frac{2\eta^2\log n}{\eps} + \frac{\eta^2\log(n)}{\eps} = \frac{3\eta^2 \log(n)}{\eps}
\end{equation}
by \Cref{eq:B-size} and \Cref{eq:cut-size}. By induction, $\cT_A$ and $\cT_O$  in line~\ref{line:cTA} and line~\ref{line:cT-o}, respectively, both have treewdith at most $8\eta^2 \log(n)/\eps$. Attaching them to $\cT$ does not increase their widths, and hence the final treewdith bound remains  $8\eta^2 \log(n)/\eps = O(\log n/\eps)$, as claimed. 
\end{proof}

\begin{remark}\label{rm:embedding} There are two places where we need planarity in the embedding theorem: (1) the existence of stochastic balanced cuts (\Cref{lm:distribution-cuts}) and (2)  the contracted graph $G^-$ remains planar so that we can invoke recursion. Since a minor-closed graph class is closed under edge contraction and vertex deletion, extending the embedding to minor-free graphs requires only showing that stochastic balanced cuts exist for minor-free graphs.
\end{remark}

\section{Stochastic Balanced Cuts}\label{sec:balanced-cuts}

This section constructs stochastic balanced cuts with low distortion in \Cref{lm:distribution-cuts}.  If we assign every edge $e\in G$ a non-negative weight $w(e)\geq 0$, then \Cref{lm:distribution-cuts} implies that there exists a cut $\cC$ such that:
\begin{equation}\label{eq:weight-cut}
     \sum_{e} w(e) (\ell(\rho_{\cC}(e))) \leq (1+\frac{O(1)}{k})\sum_e w(e)\ell(e)
\end{equation}
By Yao's minimax principle~\cite{Yao77}, the converse is true: if for any non-negative weight function $w: E\rightarrow \real_{\geq 0}$ on the edges, there exists a cut $\cC$ satisfying \Cref{eq:weight-cut}, then \Cref{lm:distribution-cuts} holds. Thus, for showing \underline{the existence} of stochastic balanced cuts, it suffices to prove the following lemma.

\begin{lemma}\label{lm:minimax-weight} Let $G = (V,E,\ell)$ be a planar graph and $B$ be a set of boundary vertices. Let $k$ be a parameter sufficiently larger than $1$ such that $|B|\leq k^2$. For any edge weight function $w: E\rightarrow \real_{\geq 0}$, one can construct in polynomial time a balanced $B$-cut $\cC$ such that:
\begin{equation*}
    \sum_{e} w(e) (\ell(\rho_{\cC}(e)) - \ell(e)) \leq \frac{O(1)}{k}\sum_e w(e)\ell(e)
\end{equation*}
and $|\cC| = O(k)$.
\end{lemma}

If we think of $\ell(\rho_{\cC}(e)) - \ell(e)$ as the detour cost for an edge $e$ cut by $\cC$, then \Cref{lm:minimax-weight} seeks a balanced $B$-cut $\cC$ such that the total weighted detour cost is small. 

In \Cref{subec:deterministic-cut}, we show how to construct a balanced cut in \Cref{lm:minimax-weight} in polynomial time. However, such a polynomial-time algorithm does not immediately imply a polynomial-time algorithm for sampling a stochastic balanced cut: the minimax principle only implies the existence of a distribution of such cuts. In \Cref{subsec:minimax}, we give a polynomial-time sampling algorithm, based on the multiplicative weight update method. Our algorithm is inspired by the algorithmic minimax for clan embeddings by Filtser and Le~\cite{FL21} (Lemma 4 in the arXiv version of their paper).

\subsection{A Cut with Small Total Weighted  Detour}\label{subec:deterministic-cut}

This section proves \Cref{lm:minimax-weight}. The proof closely follows the ideas outlined in \Cref{subsec:tech-ideas}; the main difference is how we set up the measure to capture both vertex and edge measures. 
 We will use two known tools: the rooted path separator (\Cref{lm:forest-separator-prelim}) and small hitting sets for balls in planar graphs (\Cref{thm:ball-traversal}). For rooted path separators, we need a smaller balanced constant, $1/8$ instead of $1/2$ as stated in \Cref{lm:forest-separator-prelim}; we can achieve this by applying \Cref{lm:forest-separator-prelim} a few times.

\begin{lemma}\label{lm:forest-separator} Let $G = (V,E)$ be a planar graph with a vertex weight function $\omega$, and $F$ be a rooted spanning forest that has $s$ connected components. Then there exists a balanced separator $\cP$ of $G$ w.r.t $\omega$ consisting of $O(\sqrt{s})$ rooted paths of $F$ such that for every connected component $C\in G\setminus \cP$, $\sum_{v\in C}\omega(v) \leq \omega(V)/8$  where $\omega(V) = \sum_{v\in V}\omega(v)$.
\end{lemma}

\paragraph{Setting up the proof: the measures.} Herein, we focus on proving \Cref{lm:minimax-weight}. For every vertex $v\in V$, we assign a weight:
\begin{equation}\label{vertex-weight}
   \omega_V(v) = 1/n +  \begin{cases}
    1/|B| & \text{if } v\in B  \\
    0      & \text{otherwise }
\end{cases}
\end{equation}
Let $\omega_V(V) = \sum_{v}\omega_V(v)$. Then $\omega_V(V) \in \{1,2\}$ where $\omega_V(V)= 1$ if and only if $B=\varnothing$.

We think of $G$ as a continuous graph, where every edge is a continuous segment, joined together via the incident vertices. Let \EMPH{$\cE \coloneqq \cup_{e\in E}e$}, which is the union of these continuous edges. (Some papers, e.g.,~\cite{FLMPP26}, refer to $\cE$ as a length space.)  We still refer to the points joining edges of $\cE$ as the \EMPH{vertices}.  The distance function $d_{\cE}$ for any pair of points in $\cE$ is defined as the length of the shortest path between them. At vertices of $G$, $d_G(u,v) = d_{\cE}(u,v)$ for any $u,v\in V(G)$.  

Viewing the edges of $G$ as continuous makes it easy to define measures and ultimately support a region-growing type of argument for the final proof.  We define a non-negative \EMPH{measure $\mu$} where each segment of length $\delta \subseteq e \subseteq \cE$ is assigned \EMPH{$\mu(\delta) \coloneqq w_e\cdot \delta$}. Intuitively, $\mu(\delta)$ is the weight of the $\delta$-length segment of $w_e$.  If $\delta$ is a  connected subset of $\cE$, which could span multiple edges, we define \EMPH{$\mu(\delta) \coloneqq \sum_{e\in E}\mu(\delta_e)$} where $\{\delta_e \coloneqq \delta\cap e\}_{e\in E}$. Let $\mu(E) =\sum_{e\in E}\mu(e) = \sum_{e\in E}w(e)\ell(e)$.

 For each point $p\in \cE$, we define $B_{\cE}(p,r) = \{q: d_{\cE}(p,q) \leq r\}$ to be the \EMPH{closed ball} of radius $r$ centered at $p$  , and $B_{\cE}^\circ(p,r) = B_{\cE}(p,r) \setminus \{q: d_{\cE}(p,q) = r\}$, which is an \EMPH{open ball} of radius $r$. 

We extend the vertex-weight function $\omega_V$ to every point of $\cE$ by simply assigning $\omega_V(p) = 0$ for $p\not\in \cE\setminus V$. Thus, for any connected subset $\delta \in \cE$, $\omega_V(\delta) = \sum_{v\in V(G)}\omega_V(v)(V(G)\cap \delta)$.  Finally, we define the unified measure $\zeta$: 
\begin{equation}\label{eq:zeta-def}
    \zeta(\delta) = \frac{\mu(\delta)}{\mu(E)} + \frac{\omega_V(\delta)}{\omega_V(\cE)} \qquad \forall  \text{ connected subset } \delta\subseteq \cE 
\end{equation}
\begin{observation}\label{obs:zeta-E}  $ \zeta(\cE)  = 2$. 
\end{observation}

\paragraph{The proof outline.~} The rest of the proof goes roughly as follows. First, we construct a set $Z^+$ of size $O(k^2)$ that hits a certain set of balls $\cB$ using \Cref{thm:ball-traversal}; we clarify the role of $\cB$ shortly.  Then we construct a shortest path $F_{Z^+}$ forest rooted at $Z^+$, where ties are broken consistently. Applying \Cref{lm:forest-separator}, we obtain a set $\cP$ of $O(\sqrt{Z^+}) = O(k)$ of paths rooted at $O(k)$ vertices in $Z^+$. Let $Z$ be the roots of these paths; they will be the centers of the clusters in our final cut $\cC$.  The balanced property of the rooted path separator implies that $\cC$ is also balanced. However, we need to ensure that the set of edges cut by $\cC$ has low total weighted detour.  This is where we need to design a set of balls $\cB$ such that the total $\mu$-measure of paths in the separator $\cP$ is only $O(1/k)\mu(\cE)$. Each cluster $C$ in our final cut $\cC$ is actually formed by taking each path $P_i\in \cP$  and some vertices near $P_i$. The design of $\cB$ and the construction of a cluster from each path in $\cP$ are the key technical ideas in constructing $\cC$.

Following the outline above, we now construct the set $Z^+$. For every vertex $v\in V(G)$, let:
\begin{equation}\label{eq:rho-v}
    \rho(v) = \min\{r\geq 0: \zeta(B_{\cE}(v,r))\geq 1/k^2\}
\end{equation}
to be the minimum radius such that the closed ball $B_{\cE}(v,r)$ centered at $v$ has $\zeta$-measure at least $1/k^2$. Then for every $v\in V(G)$:
\begin{equation}\label{eq:measure-Bv}
\zeta(B_{\cE}(v,\rho(v)))\geq 1/k^2 \qquad\text{while}\qquad \zeta(B^\circ_{\cE}(v,\rho(v)))\leq 1/k^2
\end{equation}

\begin{lemma}\label{lm:Z-traversal}One can construct a set $Z^+\subseteq V$ of $O(k^2)$ vertices in polynomial time  such that
\begin{equation}
    d_{G}(v,Z^+) \leq \rho(v) \qquad \forall v\in V
\end{equation}
Furthermore, $B\subseteq Z^+$.
\end{lemma}
\begin{proof} Let $\cB = \{B_{\cE}(v,\rho(v))\cap V(G): v\in V(G)\}$. Note that $B_{\cE}(v,\rho(v))\cap V(G) = B_G(v,r_v)$ for some radius $r_v\leq \rho(v)$.  Next we argue that any set of vertex-disjoint balls of $\cB$ has size at most $6k^2$. Let $\cI$ be such a set and $I$ be the set of centers of balls in $\cI$. We note that for two vertices $u\not= v \in I$, it is possible that  $B_{\cE}(v,\rho(v))\cap B_{\cE}(u,\rho(u)) \not=\varnothing$, while  $B_G(v,r_v)\cap B_G(u,r_u) = \varnothing$. However, this case only happens when there exists an edge $(x,y)$ such that two of its endpoints are in  $B_{\cE}(v,\rho(v))$ and $B_{\cE}(v,\rho(v))$, respectively, while the points internal to $(x,y)$ are shared by both balls. Since balls in $\cI$ are vertex-disjoint, we have:
\begin{equation}
    \sum_{v\in I}\zeta(B_{\cE}(v,\rho(v)))\leq 2\zeta(\cE) \leq 4
\end{equation}
by \Cref{obs:zeta-E}. However, since $\zeta(B_{\cE}(v,\rho(v)))\geq 1/k^2$ by \Cref{eq:measure-Bv}, it must be that $|I|\leq 4k^2$.

By \Cref{thm:ball-traversal}, we can construct in polynomial time a set  $Z^+\subseteq V(G)$ of size $O(k^2)$ such that $Z^+\cap B_{\cE}(v,\rho(v)) \not= \varnothing$ for every $v\in V$. This means $d_G(v,Z^+) \leq \rho(v)$ as claimed.

Since $|B|\leq k^2$, we can simply add $B$ to $Z^+$; the size of $Z^+$ remains $O(k^2)$. 

\end{proof}

Next we construct a balanced rooted-path separator. Let $Z^+$ be the set of vertices in \Cref{lm:Z-traversal}.  Let \EMPH{$F_{Z^+}$} be the shortest path forest rooted at $Z^+$ where each vertex $v$ is in the shortest path tree of the nearest vertex in $Z^+$; ties are broken consistently.   By \Cref{lm:forest-separator}, we can construct a balanced rooted-path separator \EMPH{$\cP = \{P_1,P_2,\ldots,P_{O(k)}\}$} containing $O(\sqrt{|Z^+|}) = O(k)$ shortest paths such that each connected component $K\in G\setminus \cP$ has total vertex weight $\omega_V(V(K))\leq \omega_V(V)/8 \leq 1/4$.  Let \EMPH{$Z$} be the set of roots of the paths in $\cP$.

Next, we construct a balanced cut $\cC$ containing clusters centered at $Z$. Indeed, for each path $P_i\in \cP$, let $z_i$ be its root; the corresponding cluster $C_i \in \cC$ contains $P_i$ and vertices close to $P_i$ with respect to a normalized distance function $\phi$ defined as follows.

Let $P^{\cup} = \cup_{P_i\in \cP}P_i$, viewed as a subset of $\cE$.  For every point $x\in \cE$, we define three functions:

\begin{equation}\label{eq:r-s-phi-def}
    r(x) = d_{\cE}(x,Z^+)\qquad  s(x) = d_{\cE}(x,P^{\cup}) \qquad \phi(x) =  \begin{cases}
    s(x)/r(x) & \text{if } r(x)\not= 0  \\
    +\infty      & \text{otherwise}
\end{cases} 
\end{equation}
Basically, $\phi(x)$ is the distance from $x$ to the nearest path in $\cP$, normalized by its distance to the nearest center in $Z^+$. A key property of $\phi(x)$ is that it behaves like a distance function (from $x$ to the closest path) whenever $r(x) > s(x)$, as shown in the following lemma.

\begin{lemma}\label{lm:phi-understand} Let $x$ be a point in $\cE$ such that $r(x) > s(x)$. Let $Q$ be the shortest path (w.r.t distance $d_{\cE}$) in $\cE$ from $x$ to its nearest point in $P^{\cup}$. For any $y \in Q\setminus x$:
\begin{equation}
    \phi(y) < \phi(x)
\end{equation}
\end{lemma}
\begin{proof} Observe that $s(y) = s(x) - d_{\cE}(x,y)$ and by the triangle inequality, $r(y)\geq r(x) - d_{\cE}(x,y)$. Thus:
\begin{equation*}
    \phi(y) = \frac{s(y)}{r(y)}\leq \frac{s(x) - d_{\cE}(x,y)}{r(x) - d_{\cE}(x,y)} ~ \overset{s(x) < r(x)}{<} \frac{s(x)}{r(x)}~ = \phi(x)
\end{equation*}
as claimed.
\end{proof}

Recall that every point  $x \in P^{\cup}\setminus Z$ has $\phi(x) = 0$. Now we show that a small neighborhood w.r.t $\phi$ has small total edge measure. We need this property for our final goal, which shows that cutting out a small neighborhood only incurs a low weighted total detour.  The proof uses the fact that $d_G(v,Z^+)\leq \rho(v)$ in \Cref{lm:Z-traversal}.

\begin{lemma}\label{lm:mu-U} Let $U = \{x\in \cE: \phi(x) \leq 1/4\}$. Then $\zeta(U) = O(\frac{1}{k})$ and $\mu(U) = O(\frac{1}{k})\mu(E)$. 
\end{lemma}
\begin{proof}  For every path $P_i \in \cP$ where $i\in [|\cP|]$, let $v_i$ be the non-root endpoint of $P_i$. Recall that $P_i$ is a shortest path rooted at $z_i$, and $z_i$ is closest to $v_i$ among all vertices in $Z^+$. Thus, by \Cref{lm:Z-traversal}, we have:
\begin{equation}\label{eq:Pi-vs-rho}
    \ell(P_i) = d_G(v_i, Z^+)\leq \rho(v_i)
\end{equation}

For every $x\in U$, let $P_i$ be the path in $\cP$ closest to $x$; that is, $s(x) = d_{\cE}(x,P_i)$.  We claim that:
\begin{equation}\label{eq:dist-v_i-x}
    d_{\cE}(v_i,x) < \rho(v_i)~.
\end{equation}

To see this, let $y\in P_i$ be the closest point to $x$, and $t = \ell(P_i[v_i, y])$.  Then by the triangle inequality, we have:
\begin{equation}\label{eq:dist-v_i-x-1}
     d_{\cE}(v_i,x) \leq  d_{\cE}(x,y) + \ell(P_i[v_i, y]) = s(x)+t \leq r(x)/4 +t
\end{equation}
where the last inequality is because $x\in U$ and hence $\phi(x) = s(x)/r(x) \leq 1/4$.  On the other hand, by the definition of $r(x)$ and the triangle inequality:
\begin{equation*}
   \begin{split}
        r(x) &\leq d_{\cE}(x,z_i) \leq  d_{\cE}(x,y)  + \ell(P_i[z_i, y]) \\
        &=  s(x)  + \ell(P_i) - t\\
        &\leq r(x)/4 + \rho(v_i) -t \qquad \text{(by \Cref{eq:Pi-vs-rho} and $x\in U$)}
   \end{split}
\end{equation*}
giving $3r(x)/4 + t \leq  \rho(v_i)$. By \Cref{eq:dist-v_i-x-1}, we have:
\begin{equation*}
    d_{\cE}(v_i,x) <  3r(x)/4 + t  \leq \rho(v_i)
\end{equation*}
giving \Cref{eq:dist-v_i-x}, as claimed.

  \Cref{eq:dist-v_i-x} implies that $U$ is completely covered by open balls $\{B^{\circ}(p_i,\rho(p_i))\}_{i \in [|\cP|]}$. It follows that:
  \begin{equation}
      \begin{split}
          \zeta(U) &\leq \sum_{i \in [|\cP|}\zeta(B^{\circ}(p_i,\rho(p_i)))\\
          &<   |\cP|/k^2 \qquad \text{(by \Cref{eq:measure-Bv})}\\ 
          &= O(k)/k^2 = O(\frac{1}{k})
      \end{split}
  \end{equation}
 By the definition of $\zeta$ in \Cref{eq:zeta-def}, we have:
 \begin{equation*}
     \frac{\mu(U)}{\mu(E)}\leq \zeta(U)  =  O(\frac{1}{k})
 \end{equation*}
 giving $\mu(U) = O(\frac{1}{k})\mu(E)$. 
\end{proof}

For every $\tau \in [1/8,1/4]$, we denote $A_{\tau} \coloneqq  \{v\in V: \phi(v)\leq \tau\}$. For each path $P_i\in \cP$, let:
\begin{equation}
    C^{\tau}_{i} = \{v\in A_{\tau}: P_i \text{ is closest to $v$ among paths in }\cP\}\cup \{z_i\}
\end{equation}
where ties are broken consistently. The root $z_i$ of $P_i$ is designated as the center of  $C^{\tau}_{i}$. It is possible that some paths in $\cP$ share the same root, and in this case, we merge all of the corresponding clusters into a single one.  Let $\cC_{\tau} = \{C^{\tau}_{i}\}$ be the resulting set of clusters.  In the following lemma, we show that $\cC_{\tau}$ is a $B$-lanced cut for every $\tau$. Then, to complete the proof of \Cref{lm:minimax-weight}, we show that a threshold $\tau$ chosen uniformly at random in $[1/8,1/4]$ has low weighted total detour.

\begin{lemma}\label{lm:sept-tau} For every $\tau \in [1/8/1,4]$, if $k$ is sufficiently larger than $1$, then $\cC_{\tau}$ is $B$-balanced and $|\cC_{\tau}| = O(k)$.
\end{lemma}
\begin{proof}
Recall that $Z$ is the set of roots of paths in $\cP$, which are also centers of clusters in $\cC_i$. Observe that $V(\cC_{\tau}) = A_{\tau}\cup Z$. Furthermore, $A_{\tau}\cap Z^+ = \varnothing$ since every vertex $z\in Z^+$ has $\phi(z) = +\infty$ by definition. Since $B\subseteq Z^+$ by \Cref{lm:Z-traversal}, $A_{\tau}\cap B = \varnothing$, so $\cC_{\tau}$ is $B$-nearly disjoint.

For every connected component $O$ of $G\setminus \cC_{\tau}$, since $\cP$ is balanced w.r.t $\omega_V$, by \Cref{lm:forest-separator}:
\begin{equation}
    \sum_{v\in O}\omega_V(v) \leq \omega_V(V)/8\leq 1/4
\end{equation}
since $\omega_V(V) \in \{1,2\}$. Thus, it follows that:
\begin{equation*}
   \begin{split}
       |V(O)| &\leq n (\sum_{v\in O}\omega_V(v)) \leq n/4\\
        |V(O)\cap B| &\leq |B| (\sum_{v\in O}\omega_V(v))  \leq |B|/4
   \end{split}
\end{equation*}
Finally, we show that $|V(\cC_{\tau})\setminus Z| = |A_{\tau}|\leq n/2$. Observe that $\zeta(A_{\tau}) \leq \zeta(U)$ and by \Cref{lm:mu-U}, $\zeta(A_{\tau}) = O(\frac{1}{k})$, giving:
\begin{equation}
    |A_{\tau}| \leq n(\omega_V(A_{\tau})) \leq n \zeta(A_{\tau}) \omega_V(V) \leq 2n \cdot O(\frac{1}{k})\leq n/4 
\end{equation}
as long as $k$ is sufficiently larger than $1$.  Thus, $\cC_{\tau}$ is a balanced $B$-cut.

The size $|\cC_{\tau}| = |Z|\leq |\cP| = O(k)$, as claimed.
\end{proof}

Finally, for the proof of \Cref{lm:minimax-weight}, we only need to show that choosing $\tau$ uniformly at random in $[1/8,1/4]$ gives $\cC_{\tau}$ that has low weighted total detour.

\begin{proof}[Proof of \Cref{lm:minimax-weight}]  For every $\tau\in [1/8,1/4]$. let $\rho_{\tau} \coloneqq \rho_{\cC_{\tau}}$ be the routing scheme induced by $\cC_{\tau}$.  For every edge $e = (u,v)$ cut  by $\cC_{\tau}$, if one of its endpoints, say $u\in Z$, then the detour is $\ell(\rho_{\tau}(e)) - \ell(e)  = 0$. Thus, we only need to consider the case $u\in A_{\tau}$. In this case, let $x_{\tau}$ be the point in $e$ where $\phi(x_{\tau}) = \tau$.

Let $C^{\tau}_i\in \cC_{\tau}$ be the cluster containing $u$ for some $i\in [|\cP|]$, and  $a_u$ be the vertex closest to $u$ in $P_i$. Note that $\rho_{\tau}$ routes $u$ to $z_i$, the root of $P_i$ (via a shortest path in $G[V(\cC_{\tau})]$), and then $z_i$ to $v$ (via a shortest path in $G$). By the triangle inequality:
\begin{equation}\label{eq:detour-e-1}
    \begin{split}
        \ell(\rho_{\tau}(e)) - \ell(e) & \leq d_{G[V(\cC_t)]}(u,z_i) +  d_{G}(z_i,v) - \ell(e) \\
        &\leq d_{G[V(\cC_t)]}(u,z_i) + (\ell(e) + d_{G}(u,z_i)) - \ell(e) \qquad \text{(triangle inequality)}\\
        &\leq 2d_{G[V(\cC_t)]}(u,z_i) \\
        &\leq 2(d_{G[V(\cC_t)]}[u, a_u] + \ell(P_i[a_u,z_i]))\\
        & = 2 (s(u) +  \ell(P_i[a_u,z_i]))\\
        & \leq 2 s(x_{\tau}) +  2d_{G}(a_u, z_i) \qquad \text{(since $P_i$ is a shortest path)} 
    \end{split}
\end{equation}
Next, we observe by the triangle inequality that $d_{G}(a_u, z_i) \leq s(x_{\tau}) + r(x_{\tau})$. This is because $s(x_{\tau}) + r(x_{\tau})$ is the length of the path from $a_u$ to $x_{\tau}$ and then from $x_{\tau}$ to the nearest center in $Z^+$ by definition of the function $r(x)$ in \Cref{eq:r-s-phi-def}. Plugging this bound to  \Cref{eq:detour-e-1}, we get:
\begin{equation}\label{eq:detour-e-2}
    \begin{split}
        \ell(\rho_{\tau}(e)) - \ell(e) &\leq 2s(x_{\tau})  + 2(r(x_{\tau}) + s(x_{\tau})) \leq 3r(x_{\tau})
    \end{split}
\end{equation}
since $s(x_{\tau})\leq r(x_{\tau})/4$ as $\phi(x_{\tau})\leq \tau \leq 1/4$. 

The expected error induced by $e$ by choosing random $\tau$ is then:
\begin{equation}\label{eq:expected-cut}
    \begin{split}
        \bbE_{\tau}[\rho_{\tau}(e) - e ] &\leq \frac{1}{1/4-1/8}\int_{1/8}^{1/4}  3r(x_{\tau}) d\tau\\
    &=24\int_{x_{\tau} \in e\cap \{y: 1/8\leq \phi(y)\leq 1/4\}}   r(x_{\tau}) d\phi(x_{\tau}) \qquad \text{(change of variable $\tau = \phi(x_{\tau})$)}\\
    &= 24\int_{x \in e\cap \{y: 1/8\leq \phi(y)\leq 1/4\}}   r(x) \phi'(x)dx  \qquad \text{(another change of variable $x = x_{\tau}$)}
    \end{split}
\end{equation}

Observe that $r$ and $s$ are piecewise linear on the edges of $G$ with Lipschitz constant $1$, due to the triangle inequality. Specifically, for any two points $x,y$ on the same edge $e$, if we denote $\|x-y\|_e$ to be the distance between $x$ and $y$ along the edge $e$ then:
\begin{equation*}
    \begin{split}
        |s(x) - s(y)| &\leq d_{\cE}(x,y)\leq \|x-y\|_e\\
        |r(x) - r(y)| &\leq  d_{\cE}(x,y)\leq   \|x-y\|_e
    \end{split}
\end{equation*}
Thus, it follows that, almost everywhere on the edge $e$:
\begin{equation}
    \begin{split}
        |\phi'(x)| &= |(\frac{s(x)}{r(x)}) '| = |\frac{s'(x)r(x) - r'(x)s(x))}{(r(x))^2}|\\
        & \leq \frac{r(x) + s(x)}{(r(x))^2} \qquad \text{(since $r$ and $s$ have Lipschitz constant $1$)}\\
        & \leq \frac{1 + \phi(x)}{r(x)}
    \end{split}
\end{equation}

Plugging this bound to \Cref{eq:expected-cut}, we get:
\begin{equation*}
    \begin{split}
        \bbE_{\tau}[\rho_{\tau}(e) - e ] &\leq 24 \int_{x_{\tau} \in e\cap \{y: 1/8\leq \phi(y)\leq 1/4\}}   (1 + \phi(x)) dx = 30 \ell(e\cap U) 
    \end{split}
\end{equation*}
Summing over all the edges, weighted by $w(e)$, we get:
\begin{equation*}
    \begin{split}
        \bbE_{\tau}[\sum_{e}w(e)(\rho_{\tau}(e) - e ))] &= 30\sum_{e\in E}w(e)\ell(e\cap U) \leq 30\mu(U) = \frac{O(1)}{k}\mu(E)\\  &= \frac{O(1)}{k} \sum_{e}w(e)\ell(e) \qquad \text{(by \Cref{lm:mu-U})}    \end{split}
\end{equation*}
This means there exists  a threshold $\tau \in [1/8,1/4]$ such that the total detour cost is $\frac{1}{k} \sum_{e}w(e)\ell(e)$. Such a threshold can be found in polynomial time since we only need to examine $\tau$ such that $A_{\tau}$ contains a new vertex of $V$ (compared to the previous threshold), and there are only $n$ such thresholds.
\end{proof}

The only place in the proof of \Cref{lm:minimax-weight} that requires planarity instead of minor-freeness is the existence of the rooted path separator $\cP$ of $O(k)$ paths from the rooted shortest-path forest $F_{Z+}$, whose roots are in $Z^+$. (The ball hitting set theorem holds for minor-free graphs.) Unfortunately, for minor-free graphs, such a rooted path separator does not exist due to apex graphs; see \Cref{sec:minor-free} for a more detailed technical discussion. However, the proof of \Cref{lm:minimax-weight} allows us to relax the rooted path separator condition, as stated formally in the following corollary. This relaxation turns out to be sufficient to extend the proof to minor-free graphs.

To state the corollary, we introduce the notion of $(\alpha,O(\sqrt{s}))$-path separable graphs, where one can remove $\alpha$ vertices such that the remaining graph has a rooted path balanced separator of size $O(\sqrt{s})$. More precisely, we say that $G$ is \EMPH{$(\alpha,O(\sqrt{s}))$-path separable} if for any weight function $\omega$ on the vertices of $G$, there exists a subset $X\subseteq V$ of size $|X|\leq \alpha$ such that for any rooted spanning forest $F$ with $s$ trees of $G[V\setminus X]$, there exists a balanced (vertex-weight) separator $\cP$ consisting of $O(\sqrt{s})$ rooted paths in $F$. One subtlety is that the choice of $X$ could depend on the weight function $\omega$.

\begin{corollary}\label{cor:weight-cut} Let $G = (V,E)$ be $(\alpha, O(\sqrt{s}))$-separable graph. Then \Cref{lm:minimax-weight} holds for $G$: for any parameter $k$ sufficiently larger than $1$ such that $|B|\leq k^2$ and any weight function $w: E\rightarrow \real_{\geq 0}$ on the edges, we can construct a balanced $B$-cut $\cC$ such that:
\begin{equation*}
    \sum_{e} w(e) (\ell(\rho_{\cC}(e)) - \ell(e)) \leq \frac{O(1)}{k}\sum_e w(e)\ell(e)
\end{equation*}
and $|\cC| \leq \alpha  + O(k)$. Furthermore, if we can find $X$ and the path separators in polynomial time, then we can also find $\cC$ in polynomial time.
\end{corollary}
\begin{proof} First we find $X$ of constant size as assumed by the corollary. Since $G[V\setminus X]$ is $O(\sqrt{s})$-path separable, the construction in the proof of \Cref{lm:minimax-weight} is applicable to $G[V\setminus X]$. If no connected component of $G[V\setminus X]$ has weight more than $\omega_V(V)/8$, then we simply return the cut $\cC$ containing singleton clusters: one cluster per vertex of $X$. Otherwise, there are only $O(1)$ components of weight more than $\omega_V(V)/8$; call these \EMPH{large components}.  Then we construct a balanced $(B\setminus X)$-cut $\cC_Y$ of each large connected component $Y$ in polynomial time. Let $\cC = \cup_{Y \text{ is large component}}\cC_Y$. Finally, we add $X$ to $\cC$ as singleton clusters. That is, each vertex $x\in X$ is a singleton cluster in $\cC$. The total size of $\cC$ is $\alpha  + O(k)$ since $|X|\leq \alpha$. 

For the routing cost, a key observation is that routing an edge through a singleton cluster incurs no detour cost, which gives the total weighted detour cost bound in the corollary. 
\end{proof}

\subsection{Algorithmic Minimax}\label{subsec:minimax}

As we discussed above, \Cref{lm:minimax-weight} implies that there exists a distribution $\widehat{\bbC}$ of balanced cuts that have low expected routing cost for every edge via Yao's minimax principle. However, it does not imply a polynomial-time algorithm for sampling such a balanced cut from $\widehat{\bbC}$. In this section, we give such an algorithm, based on the Multiplicative Weight Update (MWU) method. In the context of metric embedding, Filtser and Le~\cite{FL21} introduced this technique to obtain a polynomial-time samplable distribution of so-called \EMPH{clan embeddings}. Our proof closely follows theirs, except for the details of setting up the edge weight function to apply the algorithm in \Cref{lm:minimax-weight}. Using the MWU method, we obtain a set $\bbC$ of $O(n\log n)$ balanced cuts, and we sample $\bbC$ by picking a balanced cut from $\bbC$ uniformly at random.

We define an \EMPH{$\alpha$-\textsc{CutOracle}}, denoted by $\cO(G,B,\{w(e)\}_{e\in E},k)$, which takes a planar graph $G$ with $m$ edges and boundary vertex set $B$, non-negative weights on the edges, and a parameter $k$, as input. The oracle returns a $B$-balanced cut $\cC$ with $O(k)$ clusters such that the total detour cost is:
\begin{equation}\label{eq:detour-cost-total}
     \sum_{e} w(e) (\ell(\rho_{\cC}(e)) - \ell(e)) \leq \frac{\alpha}{k}\sum_e w(e)\ell(e)
\end{equation}
By \Cref{lm:minimax-weight}, we can implement  $\cO(G,B,\{w(e)\}_{e\in E},k)$ in polynomial time with $\alpha = O(1)$.

\paragraph{The algorithm.~} In the setup of the MWU method, we have a set of experts making predictions over $T$ rounds. At round $t$, each expert is given a score, and we make a ($0/1$) prediction by aggregating the predictions from all experts using their current scores. At the end of the round, the true value is revealed, and we adjust each expert's score based on whether the expert made a correct prediction.  

In our setting, each ``expert'' is an edge $e$. We use two parameters, $T$ and $\delta$, whose values will be set later in the analysis: $T$ is the number of rounds, and $\delta$ controls how each expert's score is adjusted.  In each round $t\in [T]$, we maintain a score $s_t(e)$ for edge $e$ which captures the accumulated relative detour on edge $e$. Initially we set $s_1(e) = 1$ for every edge $e$.  From the score $\{s_t(e)\}_{e\in E}$, we compute a distribution $p_t$ over the edges where:

\begin{equation}\label{eq:def-pte}
    p_t(e) = \frac{3}{4}\frac{s_t(e)}{S_t} + \frac{1}{4m} \qquad \text{where}~ S_t = \sum_{e} s_t(e)
\end{equation}

The additive term $\frac{1}{4m}$ prevents $p_t(e)$ from being too small; this way of setting the distribution $p_t$ is fairly standard.  Then we compute a weight $w_t(e) = p_{t}(e)/\ell(e)$ for each edge $e\in E$ and call the $\alpha$-\textsc{Oracle} $\cO$ to get a balanced cut $\cC_t$. 

The \EMPH{relative detour cost} of $e$ due to the routing scheme $\rho_{\cC_t}$ induced by $\cC_t$ is:
\begin{equation}
    \Delta_t(e) = \frac{\ell(\rho_{\cC_t}(e)) - \ell(e)}{\ell(e)}.~
\end{equation}
Ideally, we want this relative detour cost to be small. In the next round we update the score $s_{t+1}(e) =  s_{t}(e)(1+ \delta \Delta_t(e))$. Note that the weight $w_{t+1}(e)$ to the oracle in the next round is proportional to $s_{t+1}(e)$. Therefore, if the current detour cost $\Delta_t (e)$ is too big, $e$ will be given a larger weight  $w_{t+1}(e)$, and the $\alpha$-\textsc{Oracle} is incentivized to produce a new cut $\cC_{t+1}$ that avoids cutting $e$ to minimize the total weighted detour cost. The pseudo-code is given in \Cref{alg:MWU-cuts}.

\begin{algorithm}[H]
\caption{$\textsc{FindCuts}(G,B,k,\cO, T, \delta)$}
\label{alg:MWU-cuts}
\textbf{Input:} A planar graph $G = (V,E)$ with $n$ vertices and $m$ edges, and $B\subseteq V$ the set of boundary vertices, parameter $k$ such that $|B|\leq k^2$ and an $\alpha$-\textsc{CutOracle} $\cO$. $T$ and $\delta$ are two parameters of the WMU algorithm chosen later.\\
\textbf{Output:} a set of balanced cuts $\bbC$ of size $O(m)$. 
\begin{algorithmic}[1]
    \State $s_{1}(e)\leftarrow 1$ for every $e\in E$
    \For{$t\leftarrow 1$ to $T$}
        \State $S_t \leftarrow \sum_{e\in E}s_t(e)$
        \State $p_t(e)\leftarrow \frac{3}{4}\frac{s_t(e)}{S_t} + \frac{1}{4m}$
        \State $w_t(e)\leftarrow p_t(e)/\ell(e)$
        \State $\cC_t\leftarrow \cO(G,B, \{w_t(e)\}_{e\in E},k)$ \Comment{\lightgray{call the $\alpha$-\textsc{CutOracle} with weight $w_t(e)$ for each edge $e$}}
        \State $\Delta_t(e) \leftarrow \frac{\ell(\rho_{\cC_t}(e)) - \ell(e)}{\ell(e)}$ \label{line:deltate}\Comment{\lightgray{the relative detour cost}}
        \State $s_{t+1}(e)\leftarrow s_{t}(e)(1+ \delta \Delta_t(e))$
    \EndFor
    \State return $\bbC\coloneqq \{\cC_{t}\}_{t=1}^{T}$
\end{algorithmic}
\end{algorithm}

To show that $\cC$ is the desired distribution of balanced cuts, we follow the standard MUW analysis, using the facts in the following claim. The third item helps us upper-bound $S_{T+1}$, while the fourth item lets us show that $\delta \Delta_t(e) \leq 1$ for a sufficiently small choice of the parameter $\delta$.

\begin{claim}\label{clm:MWU-analysis} for every round $t$:
\begin{enumerate}
    \item $\sum_{e} p_t(e) = 1$ and hence $p_t$ is a distribution on the edges of $G$.
    \item  $\sum_{e\in E} w_t(e)\ell(e) = 1$.
    \item  $\sum_{e\in E} p_t(e) \Delta_t(e)\leq \frac{\alpha}{k}$.
    \item $\sum_{e\in E} \Delta_t(e)\leq \frac{4m\alpha}{k}$
\end{enumerate}
\end{claim}

\begin{proof}
 By definition of $ p_t(e)$ in \Cref{eq:def-pte}, we have:
 \begin{equation*}
         \sum_{e\in E}   p_t(e) = \sum_{e\in E} (\frac{3}{4}\frac{s_t(e)}{S_t} + \frac{1}{4m}) = \frac{3}{4}\frac{\sum_{e\in E} s_t(e)}{S_t} + \frac{m}{4m} = 3/4 + 1/4 = 1
 \end{equation*}
By definition $w_t(e) = p_t(e)/\ell(e)$. It follows that:
    \begin{equation*}
        \sum_{e\in E} w_t(e)\ell(e) = \sum_{e\in E}  p_t(e)  = 1
    \end{equation*}
Next, we have:
\begin{equation*}
    \begin{split}
        \sum_{e\in E} p_t(e) \Delta_t(e) &= \sum_{e} ( w_t(e)\ell(e)) \frac{\ell(\rho_{\cC_t}(e)) - \ell(e)}{\ell(e)}  \qquad \text{(by definition in line~\ref{line:deltate})}\\
        &= \sum_{e\in E} w_t(e) (\ell(\rho_{\cC_t}(e)) - \ell(e)) \\
        &\leq  \frac{\alpha}{k}\sum_{e\in E} w_t(e) \ell(e) \qquad \text{(by  \Cref{eq:detour-cost-total})}\\
        &= \frac{\alpha}{k} \qquad \text{(since $ \sum_{e\in E} w_t(e)\ell(e) = 1$)}.
    \end{split}
\end{equation*}
Finally, we use the fact that $p_{t}(e)\geq 1/4m$ to get:
\begin{equation*}
    \sum_{e\in E}\Delta_t(e) \leq 4m\sum_{e\in E}p_t(e)\Delta_t(e)\leq \frac{4m\alpha}{k}
\end{equation*}
as claimed.
\end{proof}

\stochasticbalancedcuts*
\begin{proof} We apply \Cref{alg:MWU-cuts} to construct a set $\bbC = \{C_t\}_{t=1}^{T}$ cuts where we set \EMPH{$T = 8m\ln(m)$}. Then we sample a balanced cut in $\bbC$ by picking a cut uniformly at random. We can sample in $O(1)$ time. For planar graphs, $m = O(n)$, so $\bbC$ has only $O(n\log n)$ cuts in the support.  Our goal is to show that:
\begin{equation*}
    \bbE_{\cC\sim \bbC}[ \frac{\ell(\rho_{\cC}(e)) - \ell(e)}{\ell(e)}] \leq \frac{4\alpha}{k} = \frac{O(1)}{k} 
\end{equation*}
since $\alpha  = O(1)$ by \Cref{lm:minimax-weight}.  By expanding the expectation and the definition of $\Delta_t(e)$ in line~\ref{line:deltate}, the above inequality is equivalent to:
\begin{equation}\label{eq:goal}
    \frac{\sum_{t\in [T]}\Delta_{t}(e)}{T} \leq \frac{4\alpha}{k}~
\end{equation}

 We follow the analysis of the MWU method: (1) upper bounding $S_{T+1}$ and lower bounding the individual score $s_{T+1}(e)$, and (2) using the fact that $S_{T+1}\geq s_{T+1}(e)$ to derive \Cref{eq:goal}. First, we bound $S_T$. Observe that:

 \begin{equation*}
     \begin{split}
         S_{t+1} &= \sum_{e\in E} s_{t+1}(e) =  \sum_{e\in E} s_{t}(e)(1+\delta \Delta_t(e))\\
         &= S_{t}  + \sum_{e\in E} \delta s_{t}(e)\Delta_t(e)\\
         &\leq  S_{t} + \delta\sum_{e\in E} (\frac{4}{3}p_t(e) S_t)\Delta_t(e) \qquad \text{(by \Cref{eq:def-pte}: $p_t(e)  \geq 3s_t(e)/(4S_t)$)}\\
         &= S_t(1 +  (4/3)\delta\sum_{e\in E}  p_t(e)\Delta_t(e))\\
         &\leq S_t(1 + \frac{4\alpha \delta}{3k}) \qquad \text{(by \Cref{clm:MWU-analysis})}
     \end{split}
 \end{equation*}

 By induction, the above inequality gives:
 \begin{equation}\label{eq:S-T-up}
     S_{T+1} \leq S_1(1 + \frac{4\alpha \delta}{3k})^{T} = m(1 + \frac{4\alpha\delta}{3k})^{T}
 \end{equation}

Next, we lower bound $s_{T+1}(e)$. Let \EMPH{$\delta = \frac{k}{4m\alpha}$}. By item 4 of \Cref{clm:MWU-analysis}, $\delta \Delta_t(e)\leq 1$. By the definition of $s_{t+1}(e)$ that
\begin{equation}
    \begin{split}\label{eq:s-t-LB}
        s_{T+1}(e)  &= s_1(e) \prod_{t\in [T]} (1  + \delta \Delta_t(e))\\
        &\geq e^{\delta\sum_{t\in [T]}\Delta_t(e)/2} \qquad \text{(use $1+x\geq e^{x/2}$ for $x\in [0,1]$)}
    \end{split}
\end{equation}
Since $S_{T+1}\geq s_{T+1}(e)$, by \Cref{eq:S-T-up} and \Cref{eq:s-t-LB}, we get:
\begin{equation}
    \begin{split}
\frac{\sum_{t\in [T]}\Delta_t(e)}{T} &\leq \frac{2}{\delta}(\frac{\ln m}{T} + \ln(1 + \frac{4\alpha\delta}{3k}))\\
    &\leq \frac{2\ln(m)}{T\delta} + \frac{8\alpha}{3k} \qquad \text{(use $\ln(1+x)\leq x$ for every $x\geq 0$)}\\
    & = \frac{8m\ln(m)\alpha}{T\cdot k} + \frac{8\alpha}{3k}  \qquad \text{(since $\delta = k/(4\alpha m)$)}\\
    &= \frac{\alpha}{k}  + \frac{8\alpha}{3k}  < \frac{4\alpha}{k}  \qquad \text{(since $T = 8m\ln(m)$)}
     \end{split} 
\end{equation}
giving \Cref{eq:goal}.
\end{proof}
\section{Extension to Minor-Free Graphs}\label{sec:minor-free}

The proof for planar graphs requires planarity to construct the rooted path separator $\cP = \{P_1,\ldots, P_{O(k)}\}$. For general minor-free graphs, such a rooted path separator might not exist. A standard counterexample is an \EMPH{apex graph}, obtained by adding a vertex, called the \EMPH{apex}, to a planar graph $G$ and connecting the apex to (one or more) vertices of $G$. Let $G^+$ be such an apex graph, with the apex being $v$. Let $T$ be a star graph centered at $v$ and containing the other vertices of $G$ as leaves. Then $T$ is a spanning tree of $G^+$. If $G$ is a square $\sqrt{n}\times \sqrt{n}$-grid, then there is no balanced separator of $G^+$ consisting of $O(1)$ rooted paths of $T$. Note that $G^+$ excludes $K_6$ as a minor. However, if $G$ excludes an apex graph as a minor, then a rooted path separator exists. That is, if we can identify a small set of apices to remove from $G$, then we can construct a rooted path separator. This is exactly the intuition behind the notion of $(\alpha, O(\sqrt{s}))$-separable graph in \Cref{cor:weight-cut}. Our goal is to show that $K_h$-minor-free graphs are $(\alpha, O(\sqrt{s}))$-separable with a constant $\alpha$.

\begin{lemma}\label{lm:minor-separable} If $G$ is a $K_h$-minor-free graph, then $G$ is $\{O_h(1), O_h(\sqrt{s})\}$-separable for any parameter $h\geq 1$ and the separator is constructible in polynomial time.
\end{lemma}

Next, we briefly argue that \Cref{lm:minor-separable} is all we need for the embedding in \Cref{thm:minor-tw}, restated below.

\minorembthm*
\begin{proof}By \Cref{rm:embedding}, for minor-free graphs, we only need to construct stochastic balanced cuts; more specifically, we need to prove \Cref{lm:distribution-cuts} for minor-free graphs. By (algorithmic) minimax in \Cref{subsec:minimax}, it suffices to construct a cut with total weighted detour cost for any weight function on the edges. Using \Cref{cor:weight-cut} and \Cref{lm:minor-separable}, we obtain the stochastic balanced cut, giving the theorem. Note that the constants in the support size of the cut distribution, as well as the cut size, also depend on $h$. 
\end{proof}

\subsection{Proof of \Cref{lm:minor-separable}}

First, we observe that if the graph is apex-minor-free, then we can construct a rooted path separator. The proof  is very similar to the proof of \Cref{lm:forest-separator} for planar graphs, except that we work with a Robertson-Seymour style decomposition~\cite{DT14} instead of a planar embedding.

\begin{restatable}{lemma}{pathsepApexMinor}\label{lm:path-sep-apex}
Let $A$ be a fixed apex graph and  $G = (V,E)$ be an $A$-minor-free graph.  Let $\omega$ be a non-negative vertex weight function, and let $F$ be a rooted spanning forest with $s$ connected components. Then there exists a balanced separator $\cP$ of $G$ w.r.t.\ $\omega$ consisting of $O_{|V(A)|}(\sqrt{s})$ rooted paths of $F$ such that for every connected component $C\in G\setminus \cP$,
\[
    \sum_{v\in C}\omega(v) \leq \omega(V)/2,
\]
Furthermore, the separator can be found in polynomial time.
\end{restatable}

The proof of \Cref{lm:minor-separable} relies on the Robertson-Seymour structure theorem. We specifically rely on the version of the theorem by ~\cite{CLPP23}, which we will state next.

\paragraph{The structure theorem.} Let $\cT$ be a tree decomposition of $G$, and for an edge $(X,Y)\in E(\cT)$ between two bags $X$ and $Y$, we call $X\cap Y$ the \EMPH{adhesion} of the edge. For a bag $X\in \cT$, we define the \EMPH{torso} of $X$, denoted by \EMPH{$\torso(X)$}, to be a graph obtained from $G[X]$ by turning the adhesion $X\cap Y$ into a clique for every neighbor $Y$ of $X$ in $\cT$. Every edge $(x,y)$ of the clique from an adhesion in $\torso(X)$ is assigned a length $\ell(x,y) = d_G(x,y)$. We use the following variant of the Robertson and Seymour decomposition by ~\cite{CLPP23}.

\begin{theorem}[Theorem 6.1~\cite{CLPP23}]\label{thm:RS-decomp} Let $G$ be a $K_h$-minor-free graph. Then there exists an apex graph $A$, and a tree decomposition $\cT$ of $G$ such that for every bag $X\in \cT$, there exists a subset $\apices(X)\subseteq X$ such that: (i) $|\apices(X)| = O_{h}(1)$ and (ii) the graph $\torso(X)\setminus \apices(X)$, obtained by removing $\apices(X)$ from $\torso(X)$, is $A$-minor-free.
\end{theorem}

We note that the adhesion of every edge in the tree decomposition in \Cref{thm:RS-decomp} has size $O_h(1)$ since the adhesion induces a clique of the torso of the edge's endpoints, and $\torso(X)$ is apex-minor-free after removing $O_h(1)$ vertices for every bag $X$.  The following lemma follows directly from \Cref{thm:RS-decomp}.

\begin{lemma}\label{lm:centroid-bag} Let $G$ be a $K_h$-minor-free graph and $\omega: V(G)\rightarrow \real_{\geq 0}$ be a vertex-weight function. There exists a set $Z\subseteq V(G)$ of size $|Z| = O_h(1)$ such that every connected component $O$ of $G\setminus Z$, either:
\begin{enumerate}
    \item $ \omega(O) \leq \omega(V)/8$, \underline{or}
    \item  $O$ has a tree decomposition $\cT_O$ and a bag $X_O\in \cT_O$ such that:
    \begin{enumerate}
        \item $\torso(X_O)$ is $A$-minor-free for an apex graph $A$ where $|V(A)| = O_h(1)$ and 
        \item   every connected component $C$ of $O\setminus X_O$ has $\omega(C)\leq \omega(V)/8$, $N_{X_O}(C)$ induces a clique in $\torso(X_O)$, and $|N_{X_O}(C)| = O_h(1)$.
    \end{enumerate}
\end{enumerate}
\end{lemma}
\begin{proof} Let $T$ be a tree isomorphic to $\cT$, where each vertex $t\in T$ corresponds to a bag $X_t\in \cT$. We root $T$ at some vertex.  For every vertex $t\in T$,  let $p_t$ be its parent in $T$, and $e_t = (t,p_t)$ be its parent edge. If $t$ is the root of $T$, then $p_t = \varnothing$, and $e_t = \varnothing$. We then assign $t$ a weight $\omega_T(t) = \sum_{v\in X_t\setminus X_{p_t}}\omega(v)$. Observe that $\omega_T(V(T)) = \omega(V)$.

Since $T$ is a tree, there exists a set $S$ of $O(1)$ vertices such that every connected component $U$ of $T\setminus S$ has weight $\omega_T(U) \leq \omega_T(V(T))/8  = \omega(V)/8$. Note that $T$ is a rooted tree.

We construct a set $Z$ in the lemma as follows. For each vertex $t\in S$ of $T$, let $e_t$ be the edge between $t$ and its parent in $T$. Then we add $\apices(X_t)$ and the adhesion of $e_t$ to $Z$. (If $S$ includes the root of $T$, then we only add the apices of the root bag of $\cT$ to $Z$.)  Since $|S|= O(1)$, by \Cref{thm:RS-decomp}, the number of vertices added to $Z$ is $O_h(1)$.

Let $E_S$ be the set of parent edges of vertices of $S$ in $T$. Let $U$ be a connected component of $T\setminus E_S$. Note that if we remove the root of $U$, then every resulting connected component has total weight at most $\omega(V)/8$. Let $\cT_{U}$ be the subtree of $\cT$ corresponding to $U$.  Then every connected component $O$ of $G\setminus Z$ has a decomposition $\cT_O$ isomorphic to a subtree of $\cT_{U}$ for some connected component $U$ of $T\setminus E_S$; the bags of $\cT_O$ might contain fewer vertices than the bags of  $\cT_{U^+}$ as some of the vertices are added to $Z$.

If $\omega_T(U) \leq \omega(V)/8$, then item 1 in the lemma holds for $O$. Otherwise, let $X_O$ be the root bag of $\cT_{O}$. Since $\apices(X_O)$ is added to $Z$ by construction, $\torso(X_O)$ is $A$-minor-free for some fixed apex graph $A$ by \Cref{thm:RS-decomp}, giving (2a). Furthermore, (2b) follows from \Cref{thm:RS-decomp} and the fact that removing the root of $U$ gives connected components of weight at most $\omega(V)/8$.
\end{proof}

We call the component $O$ of $G\setminus Z$ in item 2 of \Cref{lm:centroid-bag} a \EMPH{large component}, and the bag $X_O$ the \EMPH{important bag} of $O$. Our next goal is to show that:

\begin{lemma}\label{lm:root-sep-O} Let $F_O$ be any rooted spanning forest of large component $O$ with $s$ trees. Then there exists a balanced separator $\cP$ of $O$ consisting of $O_h(\sqrt{s})$ rooted paths of $F$.   
\end{lemma}

To this end, following~\cite{CLPP23}, we construct a graph $\what{\torso}(X_O)$ from $\torso(X_O)$ as follows.  For each connected component $C$ of $O\setminus X_O$,  we add a vertex $a_C$ and connect $a_C$ to every vertex in $N_{X_O}(C)$. We call $a_C$ the \EMPH{representative of $C$ in $X_O$}.  Recall that $N_{X_O}(C)$ induces a clique in $\torso(X_O)$ by item 2b in \Cref{lm:centroid-bag}. Then we assign a weight:
\begin{equation}\label{eq:weight_ac_1}
    \omega(a_C) = \omega(C)
\end{equation}
If there are other components of $O\setminus X_O$ sharing the same neighborhood $N_{X_O}(C)$ with a component $C$, then we only keep one vertex $a_C$ and set the weight $\omega(a_C)$ to be the total weight of all such components. Formally:
\begin{equation}\label{eq:weight_ac_2}
     \omega(a_C) = \sum_{C': N_{X_O}(C') = N_{X_O}(C)}\omega(C')
\end{equation}
Let  $\what{\torso}(X_O)$ be the resulting graph with weight $\omega$ on the vertices. Cohen-Addad, Le, Ma. Pilipczuk, and Mi. Pilipczuk showed that  $\what{\torso}(X_O)$ remains apex-minor-free.

\begin{lemma}[Lemma 6.2~\cite{CLPP23}]\label{lm:apex-minor-free}  $\what{\torso}(X_O)$ is $A$-minor-free for some apex graph $A$ with $|V(A)| =  O_h(1)$. 
\end{lemma}

Now we are ready to prove \Cref{lm:root-sep-O}.

\begin{proof}[Proof of \Cref{lm:root-sep-O}] We construct a spanning forest $\what{F}_X$ of $\what{\torso}(X_O)$ as follows. First, we remove all trees in  $\what{F}_X$ that do not contain any vertex of $X_O$. Next, 
we construct a spanning tree of  $\what{F}^{-}_X$ of $\torso(X_O)$ by contracting every  connected component of $F_O\setminus X_O$ to a vertex in $X_O$. Specifically, for every connected component $U$ of $F_O\setminus X_O$,  if $U$ does not contain the root of the connected component containing $U$ in $F_O$, then we let $u$ be the vertex in $X_O$ that is the parent of the root of $U$, and contract $U$ to $u$. Otherwise, we choose $u$ to be an arbitrary vertex of $X_O$ that is adjacent to $U$ in $F_O$, contract $U$ to $u$, and designate $u$ as the root of the contracted tree. After the contraction, $\what{F}^{-}_X$  is a rooted spanning forest of  $\torso(X_O)$. The final forest  $\what{F}_X$ is obtained from $\what{F}^-_X$ by adding $a_C$ as a leaf for each connected component $C$ of $O\setminus X_O$; that is, we pick a neighbor  $v$ of $a_C$ in  $\what{F}^{-}_X$ and add an edge $(v,a_C)$ to  $\what{F}^{-}_X$, making $v$ the parent of $a_C$ in  $\what{F}_X$. 

Observe that  $\what{F}_X$  contains at most $s$ rooted trees.   By \Cref{lm:apex-minor-free} and \Cref{lm:path-sep-apex}, we can construct a rooted path separator $\what{\cP}$ of $\what{\torso}(X_O)$  containing $O(\sqrt{s})$ rooted path.

Now we construct a rooted path forest $\cP_O$ for $O$ as follows.  For each (root-to-leaf) path $\what{P}\in \what{\cP}$, we do the following. First, if the leaf of $\what{P}$ is $a_C$ for some representative vertex $a_C$, we then add  to $\cP_O$ the rooted paths of $F_O$ containing vertices of the clique  $N_{X_O}(C)$; only $O_h(1)$ such paths are added to $\cP_O$ by item 2(b) in \Cref{lm:centroid-bag}. Next, after removing $a_C$ from $\what{P}$, let $\what{u}$ be the root of $\what{P}$ and $\what{v}$ be the leaf of $\what{P}$. If $\what{u}$ is also a root of $F_O$, then we simply add to $\cP_O$ the root-to-leaf path from $\what{v}$ to  $\what{u}$ in $F_O$. Otherwise, the root of the tree containing $\what{u}$ is outside $X_O$. Then we add two $\cP_O$ two root-to-leaf paths: one containing $\what{u}$ and the other containing $\what{v}$. In total, we add $O_h(1)$ paths to $\cP_O$ per path of $\what{\cP}$ and hence the number of paths added to $\cP_O$ remains $O_h(\sqrt{s})$. 

Since $\what{\cP}$ is balanced and $X_O$ is the important bag of $O$, $\cP_O$ is also balanced.    
\end{proof}

Now we prove \Cref{lm:minor-separable}.

\begin{proof}[Proof of \Cref{lm:minor-separable}] By \Cref{lm:centroid-bag}, we can first remove a set $Z$ of $O_h(1)$ vertices from $G$. Then for every large component $O$ of $G\setminus Z$, we construct a rooted path separator containing $O_h(\sqrt{s})$ paths. Since the number of large components is $O(1)$, the total number of paths remains $O_h(\sqrt{s})$. 
\end{proof}

\section{Applications to Designing PTASes}\label{sec:apps}

Let $\Pi$ be an optimization problem on a graph $G$ with edge lengths, where a feasible solution for $\Pi$ is a subset of the edges of $G$, and the solution value is the total length of the edges (plus the total penalty of vertices not in the solution for the prize-collecting problems).  Then, given an instance $\Pi_G$ of a problem $\Pi$ in a  planar graph $G$, we apply \Cref{thm:main} to design a randomized polynomial-time approximation scheme (PTAS) for $\Pi$ as follows:

\begin{itemize}
    \item \textbf{Step 1.} Given the accuracy parameter $\epsilon \in (0,1)$, embed $G$ into a graph $H$ with treewidth $\tau = O(\log n/\epsilon)$ and expected distortion $1+\epsilon$. Note that $V(H) = V(G)$.
    \item  \textbf{Step 2.} Solve the instance $\Pi_H$ of problem $\Pi$ on $H$ in time $2^{O(\tau)}n^{O(1)} = 2^{O(\log n /\epsilon)}n^{O(1)} = n^{O(1/\epsilon)}$ time, to get a solution $\opt_H$ of $\Pi$ on $H$.
   \item  \textbf{Step 3.} Lift $\opt_H$ to a solution $\widehat{\opt}_G$ of $\Pi$ in $G$ by replacing each edge $e= (u,v)\in \opt_H\cap E(H)$ by a shortest path $\pi_G(u,v)$ in $G$.
\end{itemize}

There are three properties that we need from $\Pi$ for the above algorithm to be applicable:
\begin{enumerate}
    \item[(P1)] $\Pi$  can be solved on treewidth-$\tau$ graphs with $n$ vertices in time $2^{O(\tau)}n^{O(1)}$.
    \item[(P2)] For the optimal solution $\opt_G$ of $\Pi_G$, (a subgraph of) $\widehat{\opt}_H = \cup_{(u,v)\in \opt_G}\pi_H(u,v)$ is a feasible solution of $\Pi_H$ in $H$.
    \item[(P3)] Compute a feasible solution $\widehat{\opt}_G$   of $\Pi_G$ in Step 3 in polynomial time.
\end{enumerate}

If $\Pi$ has the three properties above, then we show that $\Pi$ admits a PTAS in planar graphs. 

\begin{lemma}\label{lm:framework} Let $\Pi$ be a problem satisfying the three properties (P1)-(P3). Let $G = (V,E,\ell)$ be any given $K_h$-minor-free graph with $n$ vertices. Then $\Pi_G$ admits a randomized PTAS with running time $n^{O_h(1/\epsilon)}$ and expected approximation ratio $1+\epsilon$ for any $\eps\in (0,1)$.
\end{lemma}

\begin{proof} We show that the algorithm above gives the randomized PTAS.     Let $\opt_G$ be the optimal solution for $\Pi_G$. Let $\widehat{\opt}_H = \cup_{(u,v)\in \opt_G}\pi_H(u,v)$. In expectation, we have:  
    \begin{equation}
       \begin{split}
            \bbE[\ell_H(\widehat{\opt}_H)] &\leq \sum_{(u,v)\in \opt_G}\bbE[d_H(u,v)] \leq \sum_{(u,v)\in \opt_G}(1+\eps)d_G(u,v) \\&= (1+\eps)\sum_{(u,v)\in \opt_G}\ell(u,v) = (1+\eps)\ell(\opt_G).
       \end{split}
    \end{equation}

By property (P2), $\bbE[\ell_H(\opt_H)]\leq \bbE[\ell_H(\widehat{\opt}_H)]\leq  (1+\eps)\ell(\opt_G)$. By property (P1), we can find $\opt_H$ in time:
\begin{equation*}
    2^{O(\log n/\eps)}n^{O(1)} = n^{O(1/\eps)}
\end{equation*}
By property (P3) and the fact that the embedding is dominating, we have:
\begin{equation}
    \bbE[\ell(\widehat{\opt}_G)] \leq \bbE[\ell_H(\widehat{\opt}_H)] \leq  (1+\eps)\ell(\opt_G)
\end{equation}
Thus, the expected approximation ratio is $1+\eps$.
\end{proof}

There are many problems that satisfy all three properties (P1)-P(3) above, including TSP~\cite{BCKN15}, subset TSP~\cite{SS22}, Steiner tree~\cite{BCKN15}, $k$-MST~\cite{BCKN15}, and other budget variants such as $k$-TSP and $k$-Stroll, defined below.

\begin{enumerate}
    \item $k$-TSP: find a shortest closed walk visiting at least $k$ vertices in $G$. 
    \item $k$-Stroll: given a source $s$ and a sink $t$, find a shortest $s$-to-$t$ walk visiting at least $k$ vertices in $G$.
    \item Prize collecting Steiner tree: in this problem, the graph $G = (V,E,\ell)$ also has nonnegative vertex penalties $\pi:V\to\mathbb{R}_{\ge0}$. The goal is to find a Steiner tree $T\subseteq G$ minimizing the total edge length and vertex penalty:
    \begin{equation*}
        \sum_{e\in E(T)}\ell(e) +\sum_{v\in V\setminus V(T)}\pi(v).
    \end{equation*}
\end{enumerate}

Both $k$-TSP and $k$-Stroll can be solved on treewidth-$\tau$ graphs with $n$ vertices in time $2^{O(\tau)}n^{O(1)}$~\cite{SS22}. The prize-collecting Steiner tree can also be solved in $2^{O(\tau)}n^{O(1)}$ time by slightly adapting the Steiner tree algorithm~\cite{BCKN15}.

\ptases*

\bibliographystyle{alphaurl}
\bibliography{ref}

\appendix
\section{Yet Another Planar Separator Theorem}

In this section, we show that \Cref{lm:forest-separator} follows from the following separator theorem by Cai, Har-Peled, and Ye~\cite{CHY18}.

\begin{lemma}[Theorem 1.1~\cite{CHY18}]\label{lm:CHY} Let $G = (V,E)$ be a connected planar graph with $n$ vertices and $n -1 + t$ edges. Let $\omega: V\rightarrow \real_{\geq 0}$ be a non-negative weight function on $V$. Then $G$ has a $1/2$-balanced separator of size $O(\sqrt{t} + 1)$ that can be found on polynomial time. 
\end{lemma}

Now we prove \Cref{lm:forest-separator} restated below.

\forestseparator*
\begin{proof}
We triangulate $G$ by adding edges to its faces.  Let $H = G^* - \{e^*: e\in F\}$, where $e^*$ are the dual edge of $e$ and $G^*$ is the planar dual of $G$. Since $F$ is a spanning forest, $H$ is connected.  Then, by Euler formula, we have:
\begin{equation}
    |E(H)| - |V(H)|  = s-2 
\end{equation}
 By \Cref{lm:CHY}, $H$ has a $1/2$-balanced separator of size $O(\sqrt{s-1}  + 1) = O(\sqrt{s})$ for any non-negative weight function $H$. This means that $H$ has a tree decomposition $\cT_H$ of width $O(\sqrt{s})$.

For each vertex $v\in G$, denote by $P_v$ the rooted path from $v$ to the root of the tree in $F$ containing $v$. We construct a tree decomposition $\cT$ of $G$ where every bag  of $\cT$ can be covered by $O(\sqrt{s})$ rooted paths of $F$ as follows: 
 
 \begin{quote}       
       $\cT$ is isomorphic to $\cT_H$.  For every bag $Y$ of $\cT_H$, we construct a corresponding bag $B_Y$ of $\cT$ as follows.  For every vertex $f^*\in Y$ corresponding to a triangular face $f = \{x,y,z\}$ of $G$, we add to $B_Y$ vertices in the three rooted path $P_x, P_y, P_z$. Formally, $B_Y=\bigcup_{f^*\in Y}\bigcup_{x\text{ a corner of } f}V(P_x)$.
 \end{quote}

 Now we argue that $\cT$ is a valid tree decomposition of $G$. For every edge $e = (u,v)\in G$, there is a face $f$ containing both $u$ and $v$. By construction, there is a bag $Y$ containing $f^*$ in $\cT_H$ and hence the bag $B_Y\in \cT$ contains both $u$ and $v$.  

For every vertex $v\in G$, we need to show that the bags containing $v$ form a connected subtree of $\cT$. Let $D_v$ be the set of descendants of $v$ in the tree of $F$ containing $v$; note that $D_v$ also includes $v$. Let $S =\{f^*: \text{$f$ is a face of $G$ incident to a vertex in $D_v$}\}$. The observation here is that $S$ induces a connected subgraph of $H$. Let $\cT_H(S)$  be the subtree of $\cT_H$  containing all the bags that intersect $S$. Then  $\cT_H(S)$ is connected. This means $\{B_Y: Y\in \cT_H(S)\}$ is a connected subtree of $\cT$ and contains all the bags containing $v$.

Since $\cT$ is a tree decomposition of $G$, there is a bag $B$ of $\cT$ that is a $1/2$-balanced separator of $G$. Such a bag $B$ contains $O(\sqrt{s})$ rooted paths of $F$, which gives the lemma. 
\end{proof}
\end{document}